\documentclass[]{emulateapj}

\begin{document}

\title{A new raytracer for modeling AU-scale imaging of lines from protoplanetary disks}

\author{Klaus M. Pontoppidan\altaffilmark{1,2}}

\author{Rowin Meijerink\altaffilmark{1}}

\author{Cornelis P. Dullemond\altaffilmark{3}}

\author{Geoffrey A. Blake\altaffilmark{1}}

\altaffiltext{1}{California Institute of Technology, Division of Geological and Planetary Sciences, 
MS 150-21, Pasadena, CA 91125}
\altaffiltext{2}{Hubble Fellow, pontoppi@gps.caltech.edu}
\altaffiltext{3}{Max-Planck-Institut f{\"u}r Astronomie, Heidelberg, K{\"o}nigstuhl 17, 69117 Heidelberg, Germany}

\begin{abstract}
The material that formed the present-day Solar System originated in feeding zones in the inner Solar Nebula
located at distances within $\sim 20$\,AU from the
Sun, known as the {\it planet-forming zone}. Meteoritic and cometary material contain abundant evidence for the presence of a 
rich and active chemistry in the planet-forming zone during the gas-rich phase of Solar System formation. It is a natural conjecture that
analogs can be found amoung the zoo of protoplanetary disks around nearby young stars. The study of the chemistry and dynamics of planet formation 
requires: 1) tracers of dense gas at 100-1000\,K and 2) imaging capabilities of such tracers with 5-100 (0.5-20 AU) milli-arcsec resolution, 
corresponding to the planet-forming zone at the distance of the closest star-forming regions. 
Recognizing that the rich infrared (2-200\,$\mu$m) molecular spectrum recently discovered to be
common in protoplanetary disks represents such a tracer, we present a new general raytracing code, RADLite, that
is optimized for producing infrared line spectra and images from axisymmetric structures. RADLite can 
consistently deal with a wide range of velocity gradients, such
as those typical for the inner regions of protoplanetary disks. The code
is intended as a backend for chemical and excitation codes, and can rapidly produce spectra of thousands of lines for grids of models for comparison
with observations. Such radiative transfer tools will be crucial for constraining both
the structure and chemistry of planet-forming regions, including data from current infrared imaging spectrometers and extending to
the Atacama Large Millimeter Array and the next generation of Extremely Large Telescopes, the James Webb Space Telescope and beyond.
\end{abstract}


\section{Introduction}

The least processed components of the Solar System -- comets, asteroids and Kuiper Belt Objects (KBOs) -- are witnesses to
an active Solar Nebula endowed with a rich chemistry. Much of the 
formation and early evolution of the Solar System took place at small distances (perhaps within 20 AU) from the Sun where
densities were high. Only at later stages did the solar system expand to its present size (as defined by the Kuiper belt and Oort cloud), 
probably due to dynamical scattering processes involving planetesimals. It is thought that the KBOs were formed within 35 AU \citep{Levison03}, 
while the planets were all formed within 20 AU, Jupiter and Saturn within 10 AU \citep{Tsiganis05}. Oort cloud comets are thought to have formed in
the 5-10 AU region, and are composed of processed material that may have passed much closer to the Sun \citep{WoodenPPV}. 
One group of important tracers of the Solar Nebula are primitive meteorites, including carbonaceous chondrites, since
these can be brought into a laboratory setting and studied in great detail. Generally, such bodies formed in a relatively narrow range of
radii from 2-4 AU \citep{Morbidelli02}. 

While the remnants of the formation of the Solar System tell an exciting story, sometimes in great detail, 
they can never paint a full picture of planet formation in general because the evidence is ancient 
and hence often contaminated by processing and, perhaps more importantly, 
the Solar System is only one instance of a process that appears to be highly stochastic resulting in a wide range of
planetary configurations \citep{Udry07}. 
For these reasons, the region spanning $R=0.1-20$ \,AU in accretion disks around young stars
can be considered the primary target for observational studies of exo-planet formation. The part of
a protoplanetary disk within this range of radii is 
often putatively referred to as the {\it planet-forming region}.

Given the angular sizes (5-100 milli-arcseconds) and surface temperatures (100-2000\,K) 
of the planet-forming regions around T Tauri stars in the nearest star forming regions, they are
most readily traced by molecular and atomic lines in the infrared -- roughly 2-200\,$\mu$m.
Over the next decade, the mapping of the chemistry and structure of the planet-forming regions
within hundreds of nearby protoplanetary disks, using infrared gas-phase signatures, can be expected to
provide a critical complement to the study of mature planetary systems around main-sequence stars.

The first steps in this direction have already been taken, and an increasing number
of infrared gas-phase tracers have been identified. The CO fundamental ($v=1-0$) ro-vibrational band 
at 4.7\,$\mu$m is arguably the most accessible
infrared molecular band. It has been used extensively to trace gas in protoplanetary disks \citep{Najita03,Blake04,Salyk07}, 
and CO fundamental line emission from disks has in a few cases been directly imaged using 8-10 class telescopes
with adaptive optics \citep{Goto06,Pontoppidan08,vanderplas09}. Recently, {\it Spitzer} spectroscopy has demonstrated that the 10-36\,$\mu$m range of disks is
covered in molecular and atomic emission lines \citep{Carr08,Salyk08}, dominated by H$_2$O, OH, HCN, C$_2$H$_2$ and CO$_2$, 
but also including important atomic diagnostic lines such as the [NeII] transition at 12.81\,$\mu$m \citep{Pascucci07,Lahuis07,Najita09}. In the
visible, the [OI] 6300\,\AA~line has also been used as an inner disk tracer of OH \citep{Acke05,Fedele08,vanderplas08}.

Among the central goals for kinematic imaging of planet-forming zones will be 1) a search for protoplanets and their dynamical interaction with their
parent disks, 2) a search for chemical processes inferred for the Solar Nebula and 3) a general investigation of
chemistry in planet-forming regions to better understand the initial conditions for planet formation. Key questions that can be answered include:
If and how is water transported inside the snow line in protoplanetary disks, as was required to form the Earth's oceans 
\citep[see][for a review]{Encrenaz08}? Is it possible to draw direct connections between protoplanetary disks and the Solar 
Nebula chemistry inferred from 
primitive meteorites \citep{Lyons05,Smith09}? Are the inner holes and gaps observed in some protoplanetary disks associated
with planet formation \citep{Strom89,Calvet02,Lada06,Brown07}? Can protoplanets be detected dynamically through their 
interactions with the gaseous component of their parent protoplanetary disks? 
If so, what are their radial and mass distributions and how do these compare to the distributions of the
population of mature exo-planets? Massive protoplanets are expected to form their own circumplanetary accretion disks 
\citep{Canup02, Machida08, Ayliffe09} -- such as the disk that formed the Jovian moon system. 
Is it possible to detect circumplanetary disks and use their kinematic signature to confirm the presence of a protoplanet?
\begin{figure}
\centering
\includegraphics[width=8cm]{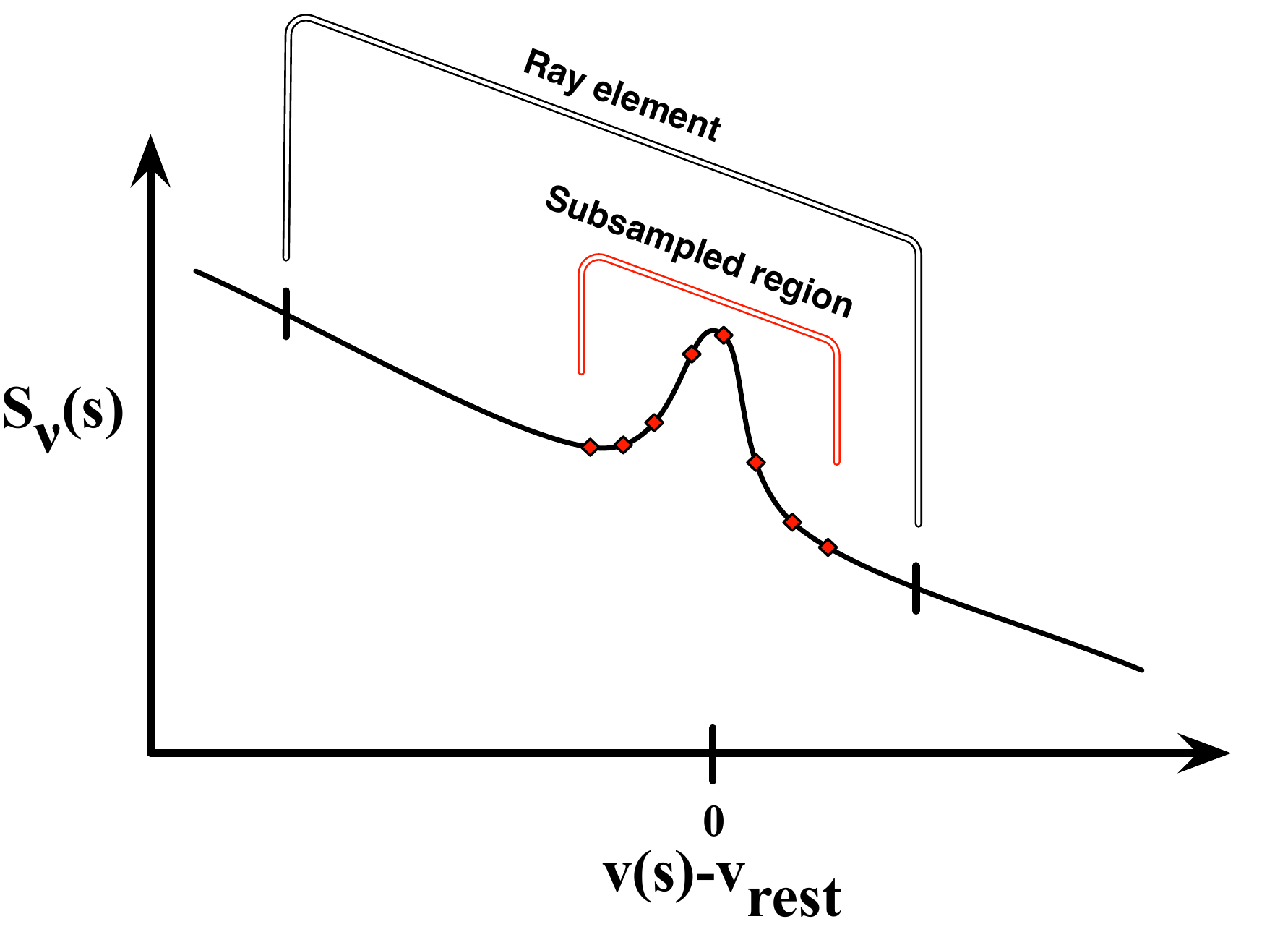}
\caption[]{Sketch of the RADLite ray element subsampling scheme. The illustration shows the line and continuum source $S_{\nu}$,as 
a function of radial velocity $v(s)$, which in turn is a function of location $s$ along the ray. Normally, the integral of the transfer equation is only
evaluated at each end of the ray element, as indicated on the figure. However, lines with narrow local broadening may be 
completely missed by the integration. In such cases, the line is localized within the ray element and the integrand evaluated in a sufficient
number of points across the line. }
\label{SC_cartoon}
\end{figure}

\begin{figure}
\centering
\includegraphics[width=9cm]{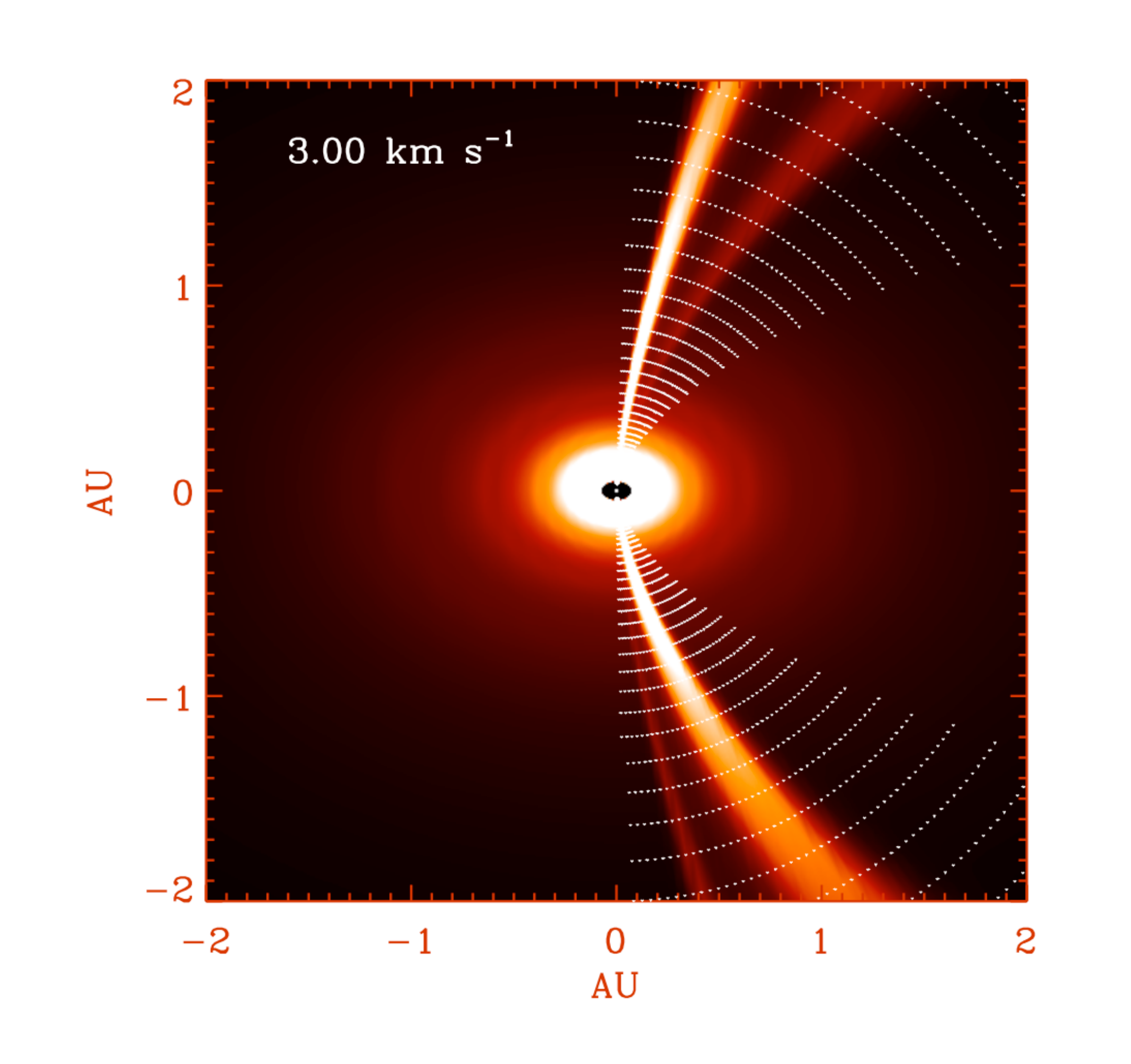}
\caption[]{Unconvolved intensity image ($I(x,y,v)$) at 3\,$\rm km\,s^{-1}$ for the 108\,$\mu$m $2_{2\ 1}\rightarrow 1_{1\ 0}$ H$_2$O line for the fiducial 
protoplanetary disk model viewed
at an inclination of 45 degrees. For illustrative purposes, the maximal water abundance is lowered to $10^{-8}$ per hydrogen nucleus to keep
the line optically thin. Due to the relatively long wavelength of this line, the disk is also has a dust optical depth of $\sim 1$, allowing
the line image to show emission from both surfaces of the disk, resulting in line emission along two isovelocity contours. 
The white points indicate rays in which RADLite has detected line interactions. The far side of the disk
is toward the top of the image.}
\label{RADLITE_rays}
\end{figure}

\begin{figure}
\centering
\includegraphics[width=8cm]{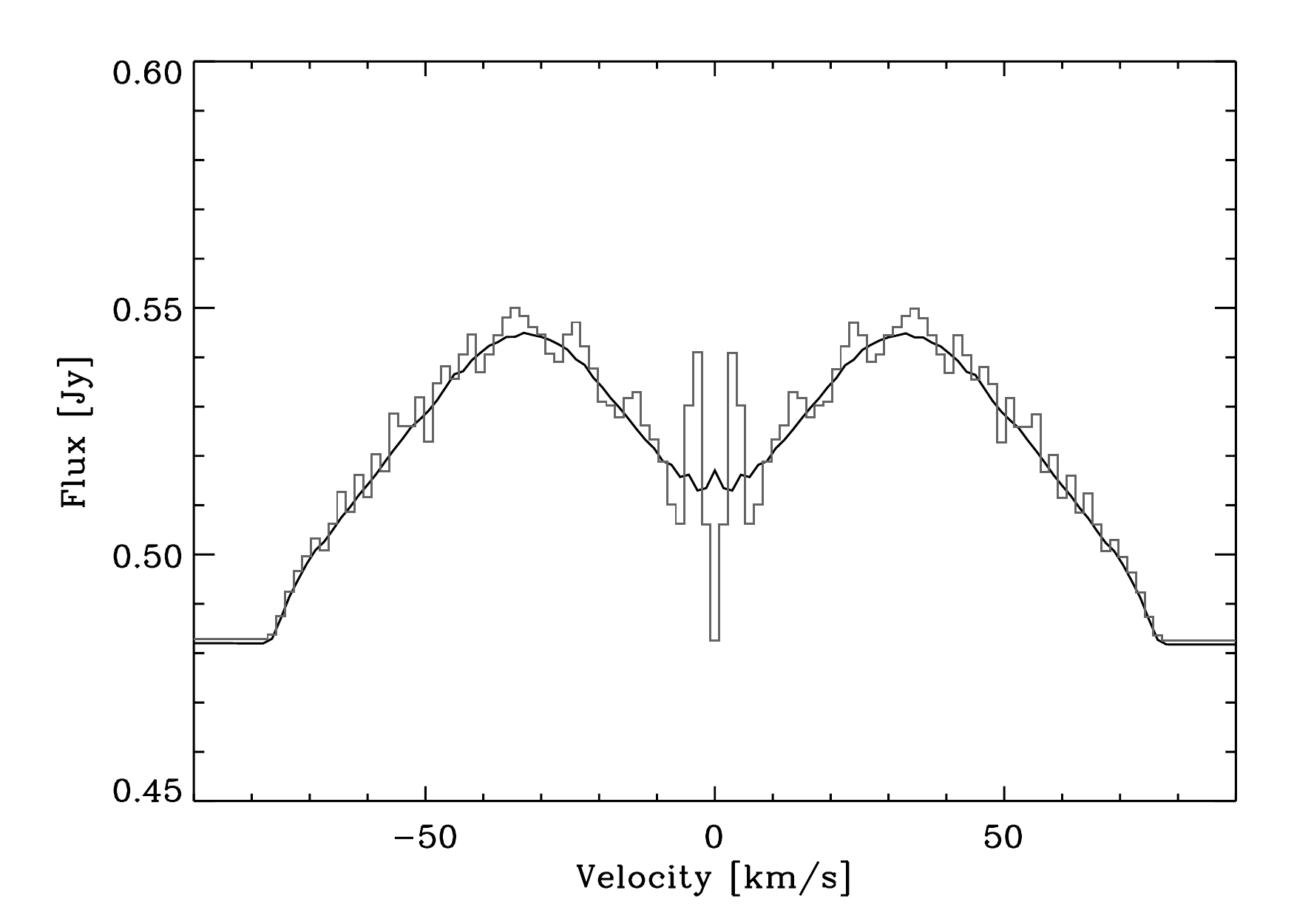}
\caption[]{Comparison of a ro-vibrational CO line rendered with (black curve) and without (grey histogram) the optimal gridding for
the same computational time. This particular model with the optimal gridding turned off could be made to produce the same result as the optimal model, 
with a choice of a very fine global grid and a large number of rays, but at the
cost of about a factor 10 in computational time. }
\label{noassist}
\end{figure}

In this paper we present a new 2D ray-tracing code, called RADLite\footnote{RADLite is currently available on a collaborative basis, with
a wider circulation anticipated in the near future.}, that is written specifically
to rapidly and efficiently render images and spectra of lines in the infrared, such as those tracing the planet-forming region,
for a full gas+dust 2D density structure. Note, however, that there is no fundamental reason why the
code cannot also be applied to lines in the visible or (sub-)millimeter region. 
Most observational studies of infrared lines use very simple models, such as slab geometries or single layer thin
disk calculations, to simulate lines from a growing database of high quality infrared line observations
\citep[e.g.,][]{Najita03,Blake04,Thi05,Brittain07,Pontoppidan08}. This is in sharp contrast to 
the relatively advanced stage of radiative transfer modeling of rotational lines in the millimeter regime \citep{Hogerheijde00,Pavlyuchenkov07}. 
With RADLite, we adopt a new philosophy: recognizing the very high complexity of inner disk chemistry and line excitation, we decouple
these steps from the simulation and fitting of observational data. Because RADLite is fast, a user is able to calculate grids of models
to determine best fitting absolute level population distributions in axisymmetric geometries. With these in hand, the application of, and comparison to, 
chemistry and excitation calculations becomes a much simpler task. An example of such an application to the mid-infrared lines of water
vapor from protoplanetary disks is presented in \cite{Meijerink09}. Here, the basic code is described in Section \ref{radlite} and the
simulation of telescopic line imaging with the future generation of Extremely Large Telescopes (ELTs) is described in Section \ref{telescope}.

As a demonstration, we apply RADLite to two examples simulating ground-based high resolution
(spectral and spatial) observations in Section \ref{Molim}: 
CO ro-vibrational emission from the fundamental ($v=1-0$) band at 4.7\,$\mu$m 
(Section \ref{COvib}) and rotational water emission lines in the N-band at 10-14\,$\mu$m (Section \ref{H2O}). 
These examples are used to demonstrate how RADLite simulates four common spectroscopic observables:
line profiles, echellograms (also known as position-velocity diagrams or 2-dimensional slit images), spectro-astrometry and integral field unit (IFU) image cubes. A
5th observable, interferometric visibilities, is not explicitly simulated, but can easily be constructed from existing RADLite products, should the need arise.
 
It is concluded, given that molecular lines in the near- to mid-infrared wavelength region are
prime tracers of the dynamics and chemistry of the planet-forming region, that the next generation
of ground-based ELTs will be ideal for producing resolved line images. This requires an
imaging spectrometer with resolving powers of $R=\lambda/\Delta\lambda =50,000 - 100,000$. 
The Atacama Large Millimeter Array (ALMA), currently under construction, is also capable of high resolution
molecular line imaging. This facility offers a tremendous improvement
over previous millimeter arrays, and has sufficient spatial resolution to provide images with 1-2\,AU resolution 
of the planet-forming region {\it in the continuum} \citep{Wolf05}. For line observations, ALMA will
reach spatial resolutions of $\sim 10-20\,$AU, or somewhat larger radii than those traced by infrared lines 
\citep{Semenov08}. Further, some molecular species, such as the symmetric rotators CH$_4$, CO$_2$ and C$_2$H$_2$, available in the infrared
cannot be reached with ALMA and vice versa. Perhaps the most notable example of a molecule generally
considered to be an infrared or space-based tracer is H$_2^{16}$O. Therefore, future observations of
infrared lines tracing the inner disk (0.1-10\,AU) will be highly complementary to the (sub-)millimeter tracers to be imaged by ALMA, and
powerful synergies between these facilities can be anticipated. 

\begin{figure}
\centering
\includegraphics[width=8cm]{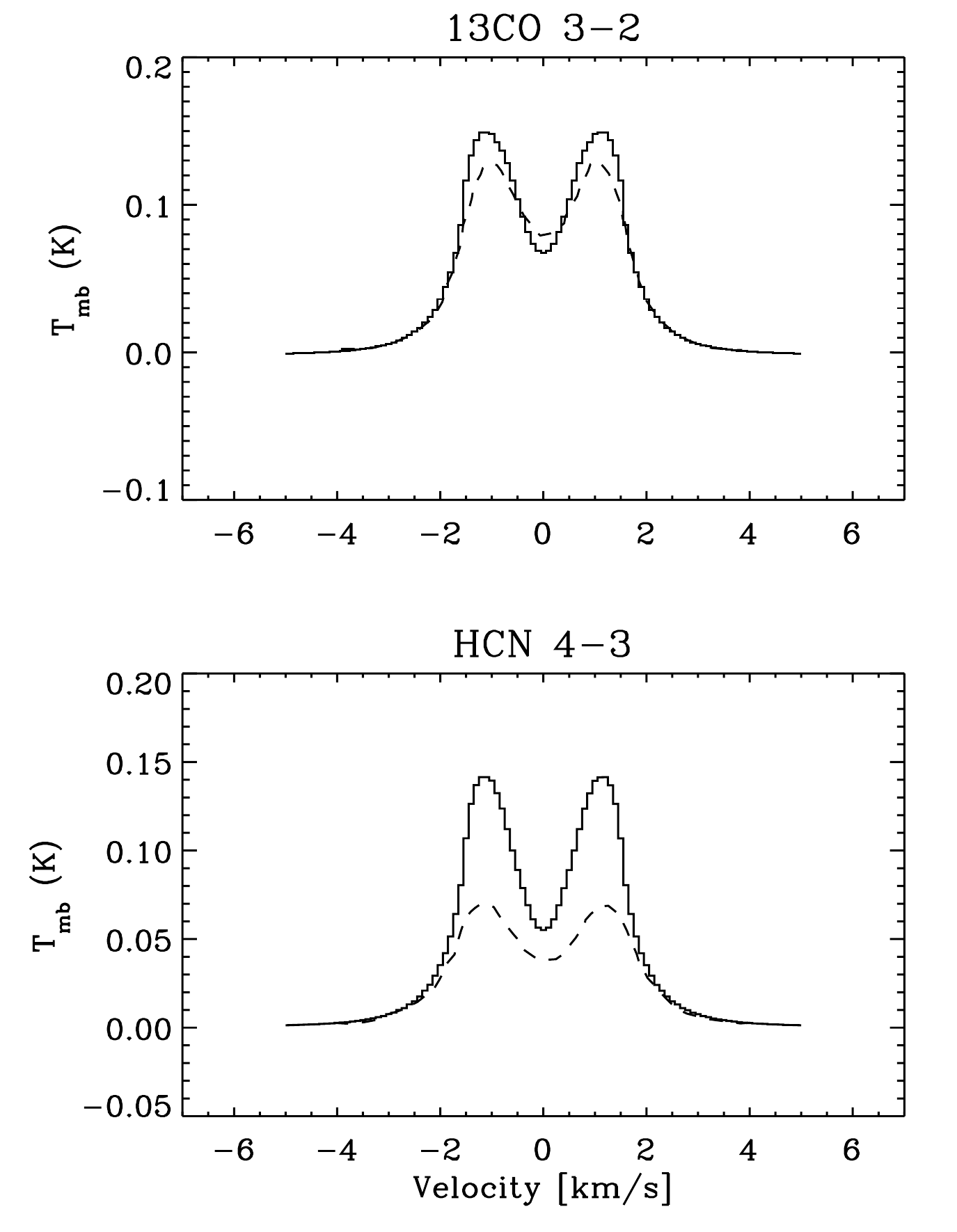}
\caption[]{Continuum-subtracted RADLite LTE renderings (solid curves) of the rotational $^{13}$CO (3-2) and HCN (4-3) lines for a disk model designed to be as
close as possible to that implemented for LkCa 15 by \cite{vanZadelhoff01} (dashed curves). The difference in strengths of the HCN lines can
be explained by subthermal excitation in the outer disk in the \cite{vanZadelhoff01} model, given the high critical density of this transition 
($1.2\times 10^8\,\rm cm^{-3}$).}
\label{lkca15}
\end{figure}

\section{Radiative transfer model}
\label{radlite}
\subsection{A raytracer for infrared lines from protoplanetary disks}
RADLite is a general ray-tracing code that has been optimized for use with infrared lines in 
axisymmetric density distributions. It is intended to be used as a back-end to
dust radiative transfer and line excitation calculations. The continuum calculation should
provide a dust temperature and the source function due to scattered photons in each grid point, as implemented 
in the continuum Monte Carlo code RADMC \citep{Dullemond04}, 
while the line excitation calculation should provide
level populations for a given molecular or atomic species. This construction is designed to take advantage
of the growing availability of multi-dimensional line radiative transfer codes that need a tool
to simulate observational data \citep[e.g.,][]{Woitke09}. In our implementation, 
RADLite uses RADMC for the dust radiative transfer, which is recommended for use with RADLite. For the line excitation 
{\it beta3D} \citep{Poelman05,Meijerink08} is used in
a 1D+1D configuration, but RADLite can interface with any axisymmetric line excitation code. 
RADLite does not assume that the 
grid spacing is regular in the radial or angular direction. If the input chemical and/or excitation
model has a fine gridding over a certain range of the angular coordinate, $\theta$, 
perhaps to adequately sample steep gradients in abundance or temperature in the surface layer of a disk, 
RADLite can simply use that grid directly. If the user does not wish to carry out a detailed
balance calculation for the level populations, RADLite can also render images and spectra assuming that the level 
populations are those given by local thermodynamic equilibrium (LTE). In all cases, RADLite accepts molecular parameter input
(levels energies, statistical weights and Einstein A values) directly from the HITRAN database \citep{Rothman05} or in the format of the Leiden Atomic and 
Molecular Database \citep[LAMBDA, ][]{Schoier05}. In the following, we use parameters from HITRAN.

\begin{figure}
\centering
\includegraphics[width=9cm]{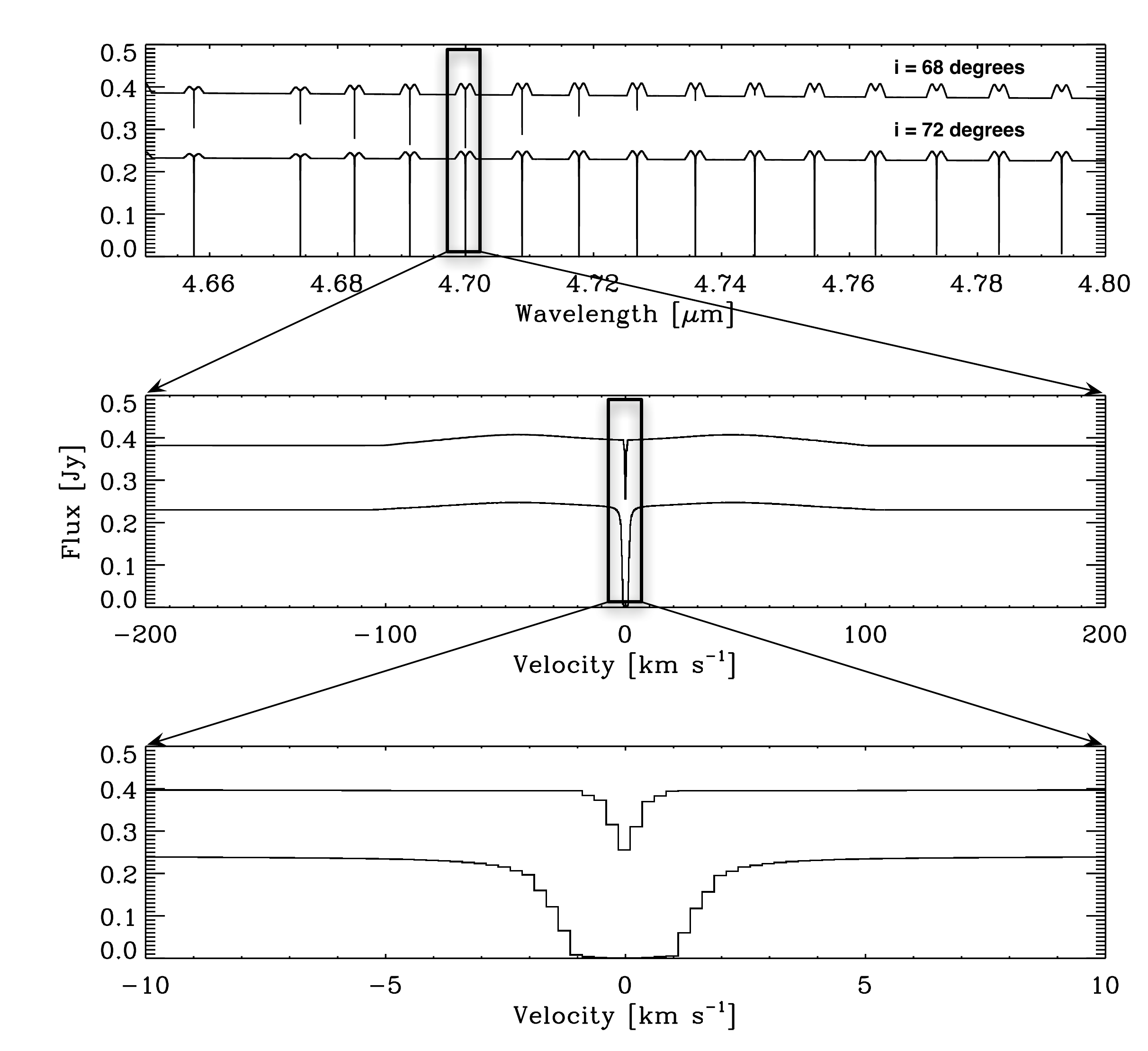}
\caption[]{Spectrum of the fundamental $^{12}$CO ro-vibrational band for a disk at inclination angles of 68 and 72 degrees,
as rendered by RADLite, showing both emission and absorption components. The spectral resolving power is assumed to be very high (0.1\,$\rm km\,s^{-1}$).
This model illustrates that RADLite can efficiently, and within one instance, deal
with spatial scales ranging from $<0.1\,$AU to $>100\,$AU as well as spectral scales ranging from $<0.1\,\rm km\,s^{-1}$ to the entire SED.}
\label{co_inclined}
\end{figure}

RADLite is based on the ``Long Characteristics'' (LC) method of integration
of the transfer equation in polar coordinates presented in \cite{Dullemond00}. LC integrations
are performed by constructing a large number of parallel rays passing through a given density structure at different 
impact parameters. Each ray is then divided into a number of ray elements defined by the 
boundaries of the density structure grid elements. The formal transfer equation is integrated across each ray element, starting from infinity. For a dust continuum
source function, this integration usually only requires that the density and temperature structures are well approximated by
a linear function across the ray element. RADLite includes the line emissivity function, $j^l_{\nu}$, and opacity, $\alpha^l_{\nu}$, terms in the 
transfer equation

\begin{equation}
\frac{dI_{\nu}}{ds} = [\alpha^c_{\nu}(R,\theta)+\alpha^l_{\nu}(R, \theta, \bar{v})][S_{\nu}(R,\theta,\bar{v})-I_{\nu}],
\end{equation}
where the source function is:

\begin{equation}
S_{\nu}=\frac{j^c_{\nu}(R,\theta)+j^l_{\nu}(R, \theta, \bar{v})}{\alpha^c_{\nu}(R,\theta)+\alpha^l_{\nu}(R, \theta, \bar{v})},
\end{equation}
and $\bar{v}$ is the macroscopic velocity projected on the ray ($R$ and $\theta$ are the polar coordinates of the grid). 

The line source and opacity functions
can vary on much smaller spatial scales than the continuum functions. Such cases typically
result from a narrow intrinsic line profile coupled with large macroscopic velocity gradients. For 
a 1-dimensional model (radial motion only) with a monotonic velocity gradient, 
the problem could be solved by applying the Sobolov, or large velocity gradient (LVG), approximation. On
a general axisymmetric grid, the LVG approximation cannot be assumed to hold, a priori, because 
a line may interact with itself at several different locations along a ray (an example
of a situation where this happens is discussed in Section \ref{COvib}). 
For an LC integration that normally evaluates the integral of the transfer equation at 
the boundaries of the model grid, a small intrinsic line broadening relative to the
global velocity gradient can result in such rapid changes of the source function that these are entirely
contained within the end points of a ray element. This situation is illustrated in Figure \ref{SC_cartoon}. 
The problem is essentially one of undersampling, and can be resolved if the disk model is
defined on a very fine grid. However, such a brute force solution generally results in a great loss of computational efficiency because
the vast majority of integration steps do not interact with a line. The result is
very long processing times -- more than 10 minutes per line on current workstations. Since an important goal of RADLite is
to be able to render grids of models, with each model potentially containing 
thousands of lines (e.g. the forest of water lines in the infrared), such inefficient raytracing 
would be unacceptable.

How bad is this undersampling problem? To explain the viscous time scales of typical PP disks, the turbulent velocities are predicted to
be small \citep{Hartmann98}. For low turbulence, intrinsic line widths are predicted to be dominated by
thermal broadening, which in some cold regions, such as the disk midplane, can be as low as tens of $\rm m\,s^{-1}$. In the inner disk, 
and for typical size grids used for the rendering of continuum images and spectra, the jump in velocities, $\Delta v$, can be
of order a few up to $10\,\rm km s^{-1}$, typically at $\sim$0.1\,AU for a T Tauri star. In other words, the length, in velocity space, of a ray element
may be 1-2 orders of magnitude larger than the width of the line. In these cases, the
contribution from the line will simply not be included in the integration, causing an erroneous 
surface brightness along that ray. RADLite solves the problem by using an optimal form of grid refinement, while still allowing the
global density structure to be defined on a relatively 
coarse spatial grid -- typically one designed for dust radiative transfer. 
For each integration step across a given ray element, the code checks for line crossings by assuming 
that the velocity projected along the ray is a smooth, linear function.  If a line crosses the ray within a ray element, 
a subdivision of the element is done in such a way that the line is
integrated with an optimal number of samples (usually about 5 per FWHM of the line). The physical parameters, density, level population, etc., are
calculated within a ray element using a simple linear interpolation between the
values evaluated where the ray crosses the cell boundary. This approach is illustrated in
Figure \ref{SC_cartoon}. 

\subsection{Ray configuration}
The optimal sampling along a ray takes care of undersampling in one dimension. 
However, proper rendering of line spectra and image cubes also requires that the set of rays is arranged
with a sufficiently fine spacing. For instance, in the case of a protoplanetary disk, infrared lines arise from a thin surface layer. This gives rise
to the line emission being formed in a thin band along the {\it isovelocity curve} in the intensity image at
a given frequence or line velocity (see Figure \ref{RADLITE_rays}). The isovelocity curve is a curve along which
the projected velocity of the emitting disk surface is constant. A disk has two surfaces, and each
surface will produce line emission along an isovelocity curve. If the disk is optically thick to continuum photons, 
only one isovelocity curve will be seen because dust blocks the observers view of the other surface. For illustrative purposes, the line in
Figure  \ref{RADLITE_rays} is chosen such that the disk has an optical depth close to unity, allowing the other surface to be seen.
The smaller the intrinsic line broadening, the narrower the emission band. If the image is not
sufficiently spatially resolved (i.e. if the pixels are not small enough), 
the thin isovelocity band may easily be missed, again resulting in severe underestimates of line fluxes. The RADLite
strategy to deal with this, without sacrificing efficiency, is to define a large number of closely spaced rays, but to only integrate those
that actually interact with the line. All other rays are considered ``continuum rays'', and are
integrated only once, assuming that the continuum emission does not change with frequency across the line. 
The rays are arranged in concentric circles with 150-600 rays per circle. The circles are spaced according to the same radial logarithmic
grid of the density structure. In the inner disk, an additional 50-100 ray circles are added spanning inwards to the
stellar surface in order to properly sample the inner rim. Line spectra are calculated by a simple surface integral; the intensities of each ray circle are averaged, 
and the total intensity calculated by a 1-dimensional radial integral of the circle averages. The star is included as a special central ray. The user can
specify whether to use a blackbody or a more complex stellar model. A typical ray configuration for optimal line rendering is shown in Figure \ref{RADLITE_rays}. 

The user can modify the sampling of the rays, depending on the application. For imaging, typically more rays are needed to avoid aliasing effects, 
i.e. artifacts caused by interpolating undersampled data from a circular grid to a rectangular grid. RADLite
outputs the circular images, i.e. the ray configuration and the associated intensities. The post-processing IDL script can then be used to 
resample the image on a rectangular mesh, with an arbitrary spatial sampling. The resampling uses TRIGRID, a bilinear interpolation 
scheme in IDL. As with all interpolation schemes, undersampling can be a problem, but the user can verify that flux is conserved by comparing
to the surface integral of the original circular image. Typical interpolation errors are of order 1\%~in the results presented here. 

\begin{figure}
\centering
\includegraphics[width=8cm]{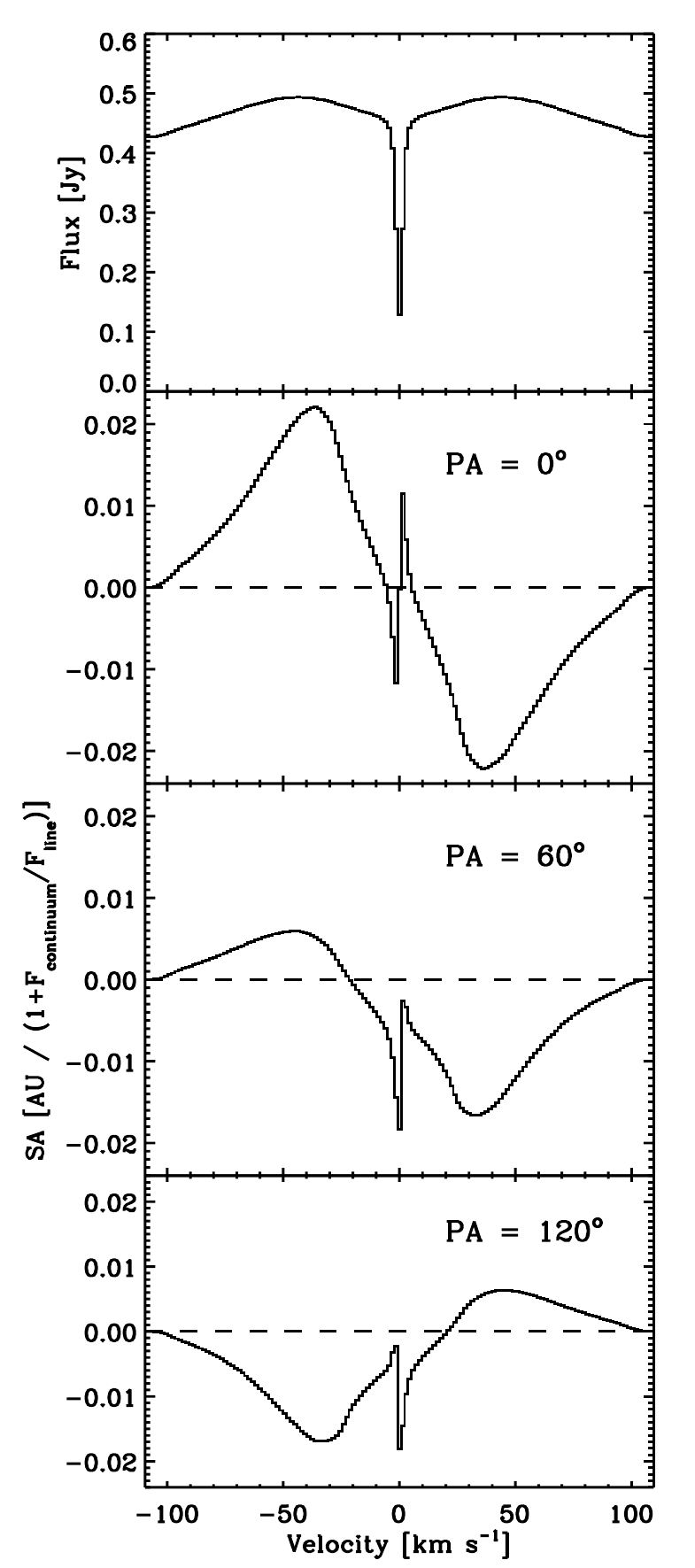}
\caption[]{Simulated astrometric spectra ($SA(v)$, Equation \ref{SA_eq}) of the CO rovibrational $v=1-0$ P(8) line for the
fiducial T Tauri star model. The data are simulated for a resolving power of 3\,$\rm km\,s^{-1}$, as is 
offered by, e.g., CRIRES on the VLT. The top panel shows the flux spectrum for a source at 125\,pc, while the lower three
panels show the astrometric spectra at three different slit position angles. 
}
\label{co_sa}
\end{figure}

Including the enhancements discussed above, RADLite renders a spectrum and image-velocity cube of a
line at 3\,$\rm km\,s^{-1}$ resolution on a single 3 GHz Intel Xeon processor core in 15-30 seconds.
Line spectra can be rendered using $\sim 50,000$ rays, while high quality images may require
up to 300,000 rays. 

\subsection{Benchmark}
There is currently no appropriate benchmark of RADLite, given the lack of other codes covering the
same parameter space. However, it was confirmed that RADLite reproduces the result of RAYTRACE, the well-tested continuum raytracer
associated with RADMC, to a fraction of a percent. This ensures that the dust radiative transfer is
carried out correctly. In Fig. \ref{noassist}, we compare a
line calculated with and without the RADLite subgridding and selective ray calculation. The
ray gridding is adjusted such that the two lines take the same number of CPU cycles to render. In practice, 
the difference between results achieved with the RADLite subgridding, as compared to brute force calculations vary
widely with line and underlying disk structure.
Furthermore, RADLite reproduces the results of the toy model fit to the SR 21 disk from \cite{Pontoppidan08} (see Section \ref{COvib}).

Finally, given that a significant volume of radiative transfer work is available for (sub)millimeter lines, 
a model was constructed to render the rotational lines of $^{13}$CO (3-2) and HCN (4-3) for a disk 
with the same parameters as those used by \cite{vanZadelhoff01} for the T Tauri star LkCa 15 based on a 
two-layer passive disk model \citep{CG97}. The main differences is that our model is based on a full 2D dust model and assumes LTE level populations, while 
the \cite{vanZadelhoff01} does not. It can be noted that the critical densities of 
the modeled lines are $\sim 3.5\times 10^{4}\,\rm cm^{-3}$ and $\sim 1.2\times 10^{8}\,\rm cm^{-3}$, so it is more likely that HCN is out of LTE than CO 
(the midplane of the \cite{vanZadelhoff01} disk has densities lower than the HCN (4-3) critical density beyond $\sim$150 AU). This is reflected
by a stronger RADLite LTE line. RADLite reproduces in detail the line strength and shape of the $^{13}$CO (3-2) transition, which is known to be in LTE throughout
the disk midplane.

\subsection{Caveats}
While RADLite is not limited to a particular geometry or velocity field, apart from the
strict requirement that the density structure is limited to 2-dimensional polar coordinates, a 
number of caveats still apply. One property of the disk emission that is not modeled in a fully self consistent manner
is scattering. Dust scattering of continuum photons is included, but is assumed to be isotropic. 
Energy emitted in the lines, however, is not able to scatter on dust. Including this would require
a Monte Carlo approach, which would be extremely time consuming -- especially if a large number of
lines are to be simulated. Lines are also rendered assuming that there are no interactions
with other lines. We find that this is generally a very good approximation for protoplanetary disks. 
In this case, line interactions will only happen when isovelocity contours overlap. For a Keplerian disk, this only occurs when
the difference in line rest frequencies between overlapping lines 
are of the order of the local line broadening or smaller, i.e. $\rm \lesssim 1\,km\,s^{-1}$. 
In the general case, for different velocity fields and molecular species, it is possible to find cases where line
overlaps are a problem. One example are the Q branches of the mid-infrared HCN, CO$_2$ and C$_2$H$_2$ bending rovibrational bands.
The shift between transitions of different isotopologues (e.g., H$_2^{16}$O to H$_2^{18}$O) is generally much too large to
cause any problems with line overlaps.

\begin{figure*}
\centering
\includegraphics[width=18cm,angle=0]{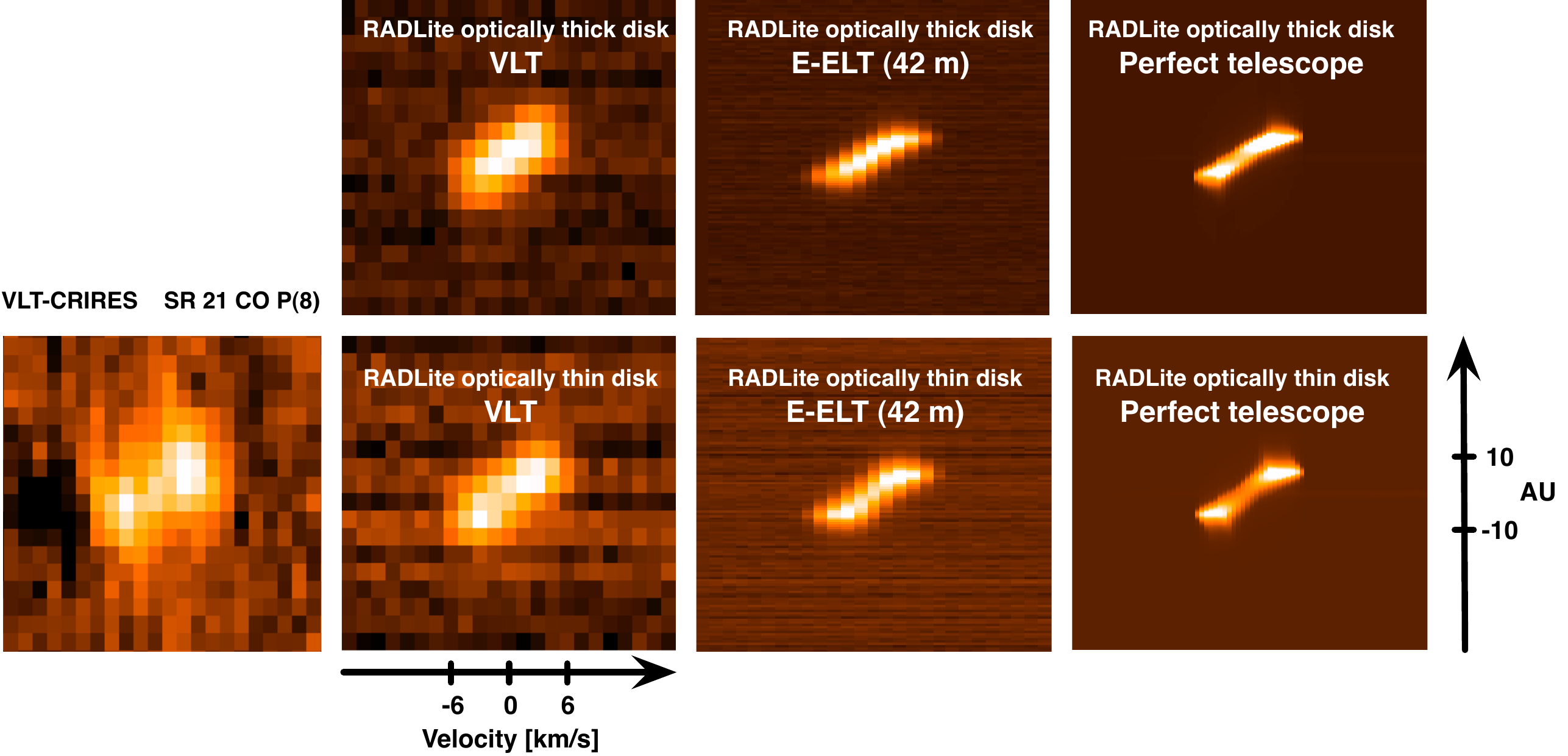}
\caption[]{Example of a VLT-CRIRES continuum-subtracted 
echellogram ($E(\eta,v)$, Equation \ref{echelle_eq}) of the P(8) CO line from the $v=1-0$ fundamental band at 4.736\,$\mu$m toward SR 21. The
line is marginally resolved in the spatial direction, with a blue-to-red peak offset of 13\,AU, as determined
using spectro-astrometry \citep{Pontoppidan08}. The right six panels show RADLite models of the line using the disk parameters
of \cite{Pontoppidan08}, i.e. a 1.5\,$M_{\odot}$, $6500\,$K, $3R_{\odot}$ star with a disk with an $R=6.5\,$AU gap, viewed at an inclination
of 20 degrees. Both the observation and model have slits oriented at an angle of 10 degrees from the disk major axis.
Otherwise, the disk is the fiducial model, with a CO abundance of $2\times 10^{-4}$ per hydrogen nucleus. The top panels is a model 
that assumes a gas-to-dust ratio of 1280 and is optically thick in the continuum at 4.7\,$\mu$m, the lower panels show a disk where the gas-to-dust 
ratio has been increased to $2.6\times 10^5$, at which point the disk becomes optically thin to continuum photons at 4.7\,$\mu$m (see the discussion in the text). 
The resolving power is 3\,$\rm km\,s^{-1}$ for both models and observation. The horizontal noise streaks is
an artifact of the continuum subtraction due to the noise of the continuum frame being reproduced at each velocity. Hence, the streaks would also appear in real data. 
Note that the models assume diffraction limited imaging (FWHM = 122 milli-arcsec at 4.7\,$\mu$m), while the actual CRIRES observation probably has a significant
seeing halo from incomplete AO correction. The point source sensitivity for the ELT observation is assumed to be 
$1.3\times 10^{-4}$\,Jy rms for a 1 hour observation, which is about 20 times better than that of VLT-CRIRES. }
\label{SR21}
\end{figure*}

\section{Simulating observations}
\label{telescope}
The primary purpose of RADLite is to simulate observations, and the package is therefore equipped with simple post-processing scripts
that can, given parameters for a virtual telescope, convert ideal images and spectra to three typical observational products: single-slit spectra, astrometric spectra, and
image cubes (the last as would be observed with an idealized IFU). However, it is
foreseen that users typically will want to supply their own telescopic models to suit some required level of complexity of the telescope simulation. 
For this reason, this step is not considered an integral part of RADLite, but is provided as an optional post-processing IDL script. In the following, the
telescope simulation that is currently used is presented. Tools that are not yet developed include
visibility calculations for current and future infrared interferometers, such as VLT-MATISSE \citep{Lopez08}. 

\subsection{Telescopic observables}

First, the way RADLite treats the basic observables; image cubes, echellograms, line profiles and spectro-astrometry is described. 
The simulations assume that observations are made with grating spectrometers (as is the case for almost all current mid-infrared astronomical instrumentation). 
A grating spectrometer images a spectrally dispersed source on a 2-dimensional detector after the image of the source (a PSF for a point source) has been truncated by passage
through a (approximately rectangular) slit with a length and a width, measured in angular units. This configuration results in a spectral
image or echellogram that has a spatial dimension (along the slit) and a spectral dimension (across the slit). Some times more than one spectral order is imaged on
the same detector in a so-called cross-dispersed configuration. It is also possible to place a number of slits next to each other and redirect the
resulting spectra to different locations on the same detector. Such an instrument is known as a slicer-type IFU, and will produce
a line image cube in which each pixel contains a spectrum (a so-called spaxel). Other types of IFUs exist but, 
for this paper, only a slicer is considered. A slicer IFU data cube can also be constructed using a single-slit 
instrument by stepping the slit across the source, although this approach obviously results in a great loss in efficiency as compared to an IFU. 

The simulated observables are calculated as follows. The basic RADLite output is an intensity map as a function of velocity offset, $v$, from a line center, 
$I(x,y,v)$, where the spatial coordinates, $(x,y)$, are measured in angular units, assuming some distance to the source. A telescopic observation causes the 
intensity map to be convolved with an image point spread function, $PSF(x,y)$, and an instrumental line broadening profile, $L(v)$, resulting in an ``telescopic'' 
intensity map:

\begin{equation}
  I_T(x,y,v) = [I(x,y,v)*PSF(x,y)] * L(v),
\label{IT_eq}
\end{equation}
where $*$ denotes convolution. 

The echellogram, as observed by a single slit grating spectrometer, is simply the telescopic intensity map integrated over the entrance slit width. The slit 
is oriented at a position angle $PA$ on the sky 
(negative rotation), with width, $W$, such that the coordinates defined by the slit become, $\xi = x\cos(PA) - y\sin(PA)$ and $\eta = x\sin(PA)+y\cos(PA)$. 
In this system $\xi$ and $\eta$ are coordinates across and along the slit, respectively. The echellogram becomes:

\begin{equation}
  E(\eta,v) = \int_{-W/2}^{W/2}I_T(\eta,\xi,v){\rm d}\xi.
  \label{echelle_eq}
\end{equation}
The advantage of the echellogram is that it encompasses both spatial and spectral information.

The line profile is a higher level product, being the echellogram integrated over some chosen aperture along the slit ($\eta$) with width, $A$. Often this integral is 
weighted by the signal strength in the $\eta$ direction, resulting in a so-called optimal extraction. However, since that approach can be difficult to quantify for extended 
sources, it is not used here. The line profile is:

\begin{equation}
  F(v) = D_A\times \int_{-A/2}^{A/2}E(\eta,v) {\rm d}\xi,
  \label{FL_eq}
\end{equation}
where $D_A$ is a constant factor, depending on the shape of the PSF, of order unity taking into account that $A$ may not be large enough to contain all the source flux.
If there is no noise added to the model, $A$ can be chosen sufficiently large so that $D_A=1$. 
It is possible to also use the echellogram to calculate the astrometric (first moment) spectrum. Following \cite{Pontoppidan08}:

\begin{equation}
  SA(v) = C_A\times \frac{\int_{-A/2}^{A/2}E(\eta,v)\eta{\rm d}\eta}{\int_{-A/2}^{A/2}E(\eta,v){\rm d}\eta},
  \label{SA_eq}
\end{equation}
where $C_A$ is a constant factor that approaches unity as $A\rightarrow \infty$, similar to $D_A$. This spectrum 
is used to extract spatial information on scales smaller than the formal spatial resolution of the telescopic image.
Note that if there is a strong continuum underlying the line, the astrometric spectrum of the line will be ``diluted'' by a factor depending
on the line-to-continuum ratio: $(1+F_{\rm continuum}/F_{\rm line})$. In other words, the measured astrometric spectrum as given in Equation \ref{SA_eq} 
should be multiplied by this factor to determine the real, continuum-subtracted, extent of the line emission.

\begin{figure*}
\centering
\includegraphics[width=16cm]{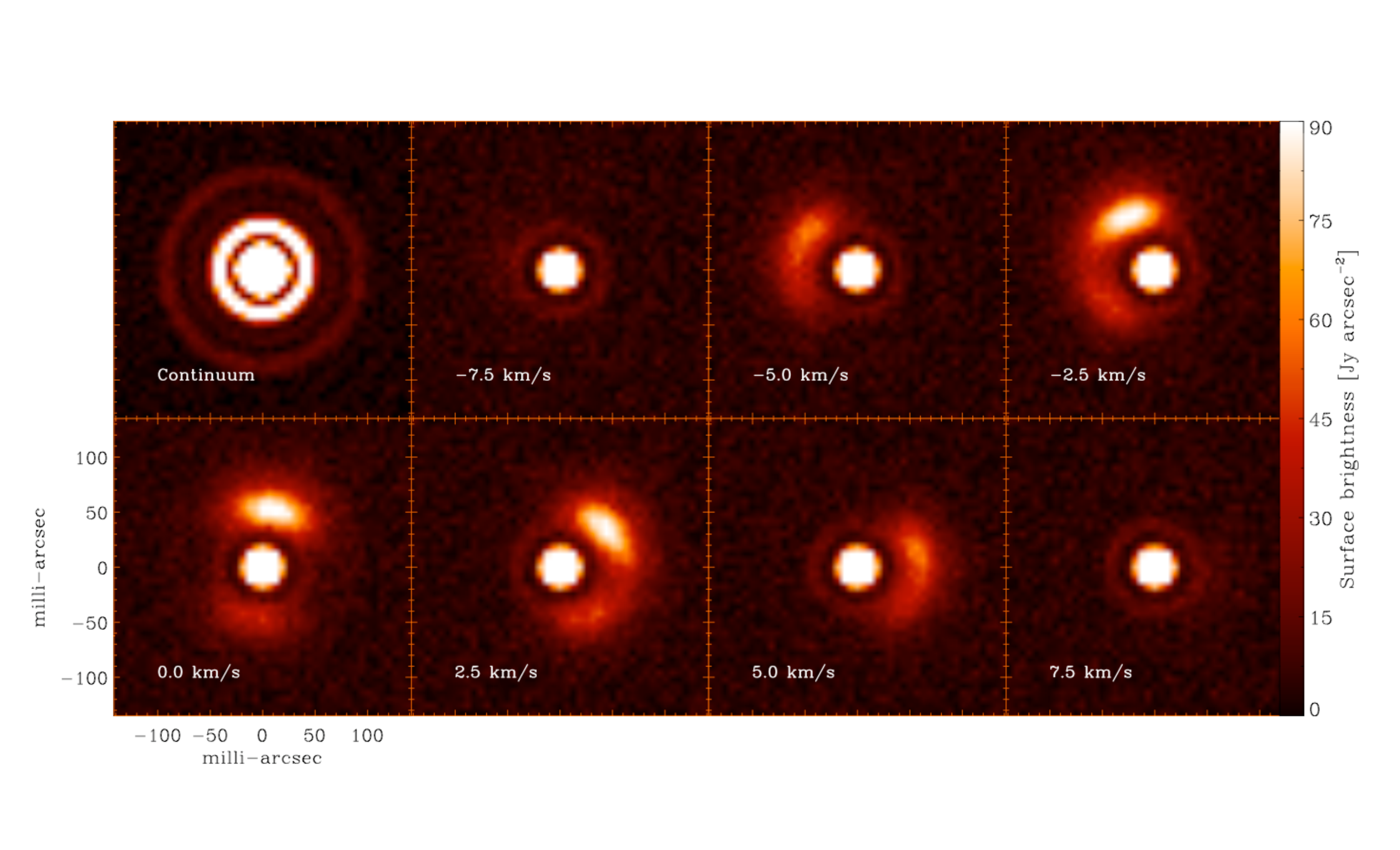}
\caption[]{Image cube ($I_T(x,y,v)$, Equation \ref{IT_eq}) of a simulated E-ELT IFU observation of the $v=1-0$ P(8) CO line at 4.736\,$\mu$m for SR 21 at 
an assumed distance of 125 pc, including
an appropriate noise level, as discussed in the text. The far side of the disk is toward the top. The scale of the image is linear.  
This simulation can be compared to the simulated single slit observation in Figure \ref{SR21}. The first panel shows the continuum image obtained bluewards of
the line. 90\% of the continuum has been subtracted from the remaining panels to illustrate that the minimum required differential imaging contrast is only 1:10. }
\label{ELT_CO}
\end{figure*}

\subsection{PSF modeling}
In order to estimate the quality of telescopic line images, an idealized diffraction PSF of an ELT is calculated and used to
convolve model images. We use the analytical approach of \citep{Sabatke05} relevant for pupil masks consisting
of hexagonal segments. Note that we do not attempt to accurately model an actual telescope, since the various ELTs are still
in the design phase.  In particular, the influence on the pupil mask by the presence of secondary 
mounts has been ignored, and the segment alignment is assumed to be perfect. The adopted ELT mirror is 
inspired by that planned for the European ELT (E-ELT), and consists of 984 hexagonal segments, each 
1.4\,m wide, resulting in a 42\,m aperture. 

\subsection{Atmospheric transmission and correction for telluric lines}

A fundamental problem for ground-based infrared spectroscopy is the presence of strong absorption lines
from molecules in the Earth's atmosphere (telluric lines). These lines are often the same transitions that may be targeted in
astronomical objects, and protoplanetary disks in particular. If the telluric line is saturated, it is necessary
to observe the target at a time of the year when its radial velocity is large enough to shift the line away from its 
telluric counterpart. Some astronomical objects have large intrinsic radial
velocities, but most nearby protoplanetary disks do not. 
Typically, shifts of order 20-30\,$\rm km\,s^{-1}$ are required, which, by a coincidence
of nature, are similar to the velocity of the Earth relative to the Sun. 
To simulate these circumstances, we use a transmission model calculated for Paranal in Chile, 
the location of the European Very Large Telescope, and assume an optimal, but generally 
achievable, velocity shift of 30\,$\rm km\,s^{-1}$. 

Space-based infrared observatories, such as the James Webb Space Telescope (JWST), obviously do not require the correction
for telluric absorption. The Stratospheric Observatory For Infrared Astronomy (SOFIA) has a significant atmospheric
column above it for terrestrially abundant species such as CO$_2$, CH$_4$, or H$_2$O, and hence still requires a correction for the telluric transmission spectrum. 

\subsection{Sensitivity}
An important component for predicting telescope performance is the sensitivity of the modeled observation. 
For a future ELT, predicting detailed sensitivities is clearly a futile task. 
For illustrative purposes, we use point source sensitivities of 125\,$\mu$Jy (1$\sigma$ in one hour) at M-band (4.6\,$\mu$m) 
and 370\,$\mu$Jy at N-band (10.5\,$\mu$m) for resolving powers of 100,000 and 50,000, respectively. These are values calculated using the online
exposure time calculator for the E-ELT, available on the ESO website. The M-band sensitivity is only $\sim 20$ times better 
than that of VLT-CRIRES, which seems to be a conservative estimate. For instance, on the website for METIS, the 
high resolution mid-infrared spectrometer E-ELT concept, the point source sensitivities are reported to be $\sim$10 times better at 4.6\,$\mu$m.  
At N-band the gain seems to be significantly higher compared to VLT-VISIR, according to all available estimates. Sensitivities for
existing instrumentation are based on actual typical performance. 

\section{Molecular line imaging in the infrared}
\label{Molim}
A key property of ground-based telescopes with extremely large apertures is their potential for delivering imaging
with very high spatial resolution in the infrared -- where the performance of adaptive optics systems is expected
to deliver nearly diffraction limited images with a resolution as high as 10-25 mas at 2-5\,$\mu$m and 50-100 mas at 10-20\,$\mu$m. 
This corresponds to physical scales of 2-10\,AU in nearby star forming regions, which offer up to 1,000 potential targets including 
PP disks and other types of young stellar objects (YSOs) across the stellar mass range from the brown dwarf limit to young A stars 
\citep{Evans09}. In this section, we demonstrate how RADLite can be applied
to existing molecular line spectra and explore what images of PP disks in various infrared molecular lines are expected 
to look like with a mid-infrared imaging spectrometer operating with spectral resolving powers of 50,000-100,000 mounted on an ELT. 
Several concepts for such ELT instruments are currently being developed. While the focus of the RADLite demonstration 
is on line imaging with ground-based facilities, there are other important facilities for mid-infrared spectroscopy, each covering
unique aspects. The mid-infrared spectrometer on the JWST, MIRI, will provide highly sensitive medium resolution spectroscopy
across the mid-infrared range (5-27\,$\mu$m). For the brightest sources, the high resolution spectrometer on SOFIA, EXES, will enable spectrally resolved observations
of a wide range of lines, not available from the ground, including the bending mode lines of water around 6\,$\mu$m, in addition
to other important species such as CH$_4$. RADLite is also intended to aid in the modeling of observations from these facilities. 

\subsection{Fiducial disk model}
We base most of the examples on the {\it fiducial disk model} described in detail in a companion paper \citep[model 5 of][]{Meijerink09}, but note
the cases where departures from the fiducial model are made for illustrative purposes, such as introducing an inner hole to simulate
transition disks. Briefly, the central source is 
a $M=1\,M_{\odot}$, $R=2\,R_{\odot}$ star with an effective temperature of $T=4275\,$K, corresponding to a luminosity
of 1.4\,$L_{\odot}$ and an age of 2 Myr \citep{Siess00}. The stellar input spectrum 
is a model atmosphere from \cite{Kurucz93}. The disk has a gas mass of $10^{-2}\,M_{\odot}$, outer radius of 120\,AU, 
a Gaussian vertical density profile and a surface density that varies with radius as $\Sigma\sim R^{-1}$. 
The inner rim is roughly located at the dust sublimation radius corresponding to 1600\,K. The disk is assumed to have already experienced
some grain growth, simulated by an enhanced gas-to-dust ratio of 12800, a low flaring index of $H/R \sim R^{-0.1}$, and outer scale height 
$H_{\rm out}/R_{\rm out} = 0.1$. The total gas-to-solid ratio will of course still be $\sim$100; 
the idea is to model the case where 99\% of the dust has already grown to centimeter or meter sizes,
at which point they no longer contribute significicantly to the
infrared opacity. Further, these large grains will likely have settled to the mid-plane, and will therefore
no longer contribute to the disk surface properties. The local line broadening is assumed to be 
dominated by the thermal component, corresponding to a low turbulent broadening, 
assumed to be a constant fraction of the local sound speed: $v_{\rm turb} = 0.03\,c_s$. 
The CO abundance is a constant $2\times 10^{-4}$ per hydrogen nucleus. 
The maximal water abundance (gas + ice) is taken to be high, $3\times 10^{-4}$ per hydrogen nucleus (i.e. H+$\rm 2H_2$)
in shielded regions. As discussed in \cite{Meijerink09}, the innermost disk has the maximal water abundance, while
the surface outwards of the mid-plane snow line, located at 1.7\,AU, is strongly depleted in water down to 
a fractional abundance of $10^{-10}$ per hydrogen nucleus in order to match observed mid-infrared water spectra in T Tauri disks. 
The level populations of the water vapor are calculated in non-LTE using beta3D, while the CO is assumed to be in
LTE. A complete, quantitative description of the RADLite input model, such as the density, temperature and abundance distributions, as
well as a detailed treatment of the excitation of water vapor, can be
found in \cite{Meijerink09}.

\begin{figure}
\centering
\includegraphics[width=9cm]{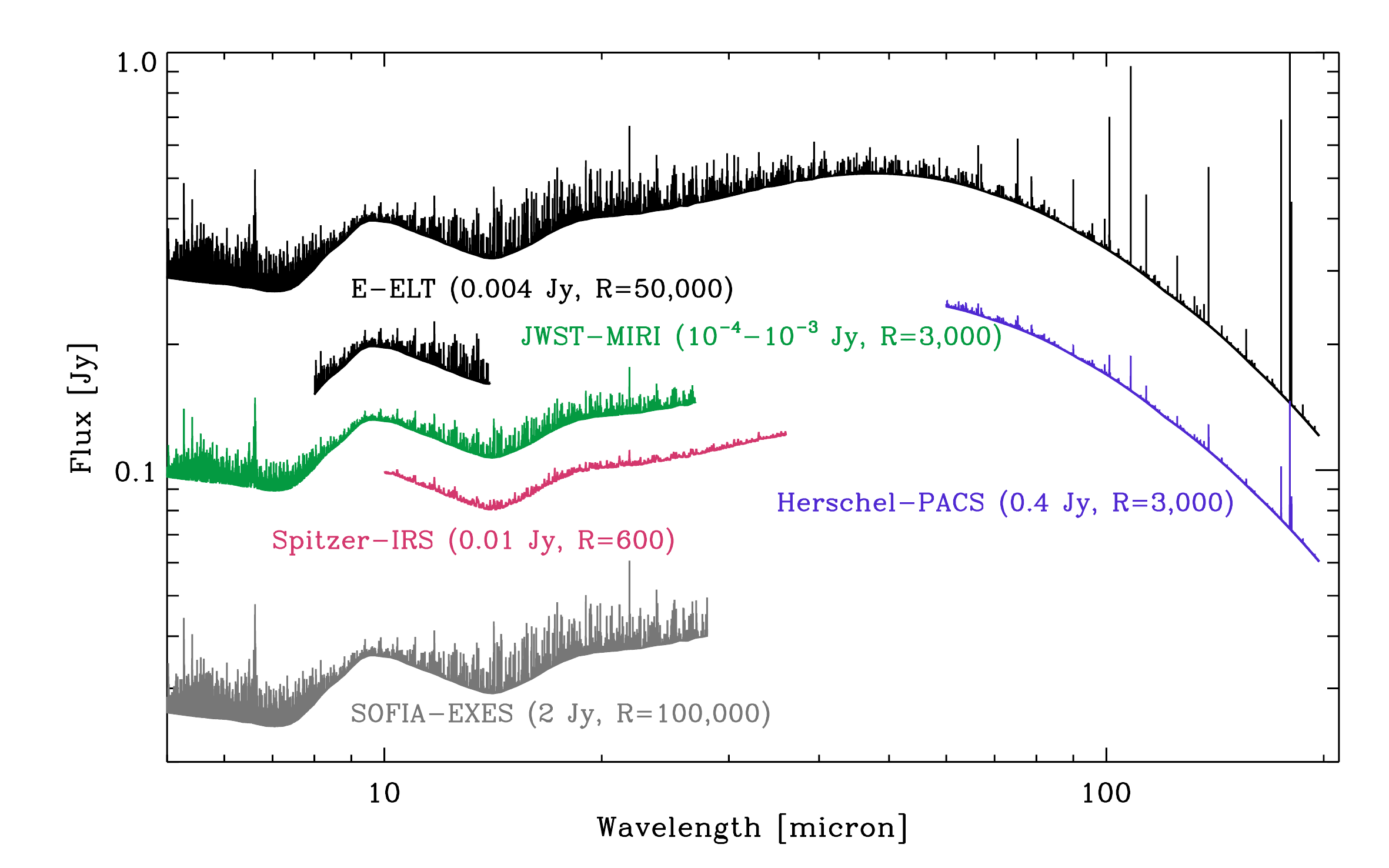}
\includegraphics[width=9cm]{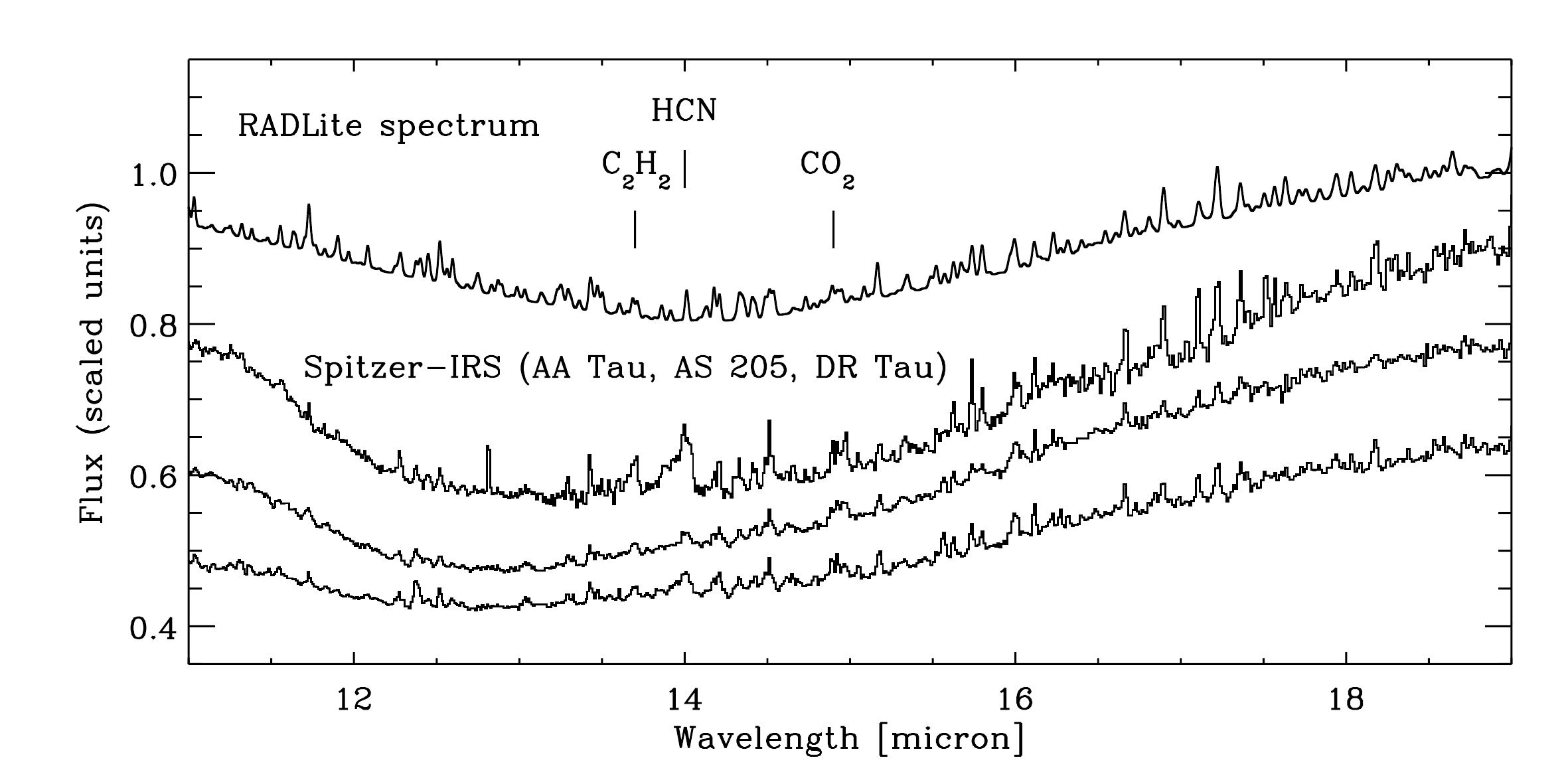}
\caption[]{Top panel: A full range calculation of the infrared water spectrum from a typical protoplanetary disk 
(model 5 in \cite{Meijerink09}). In addition, scaled (for clarity) model spectra convolved to the
spectral resolution of various infrared observatories are shown. For each observatory and instrument, rough sensitivities ($\sim 10\sigma$ in 1 hour) and
spectral resolving powers are indicated. Bottom panel: A comparison of the model convolved to the Spitzer resolution (top curve) with {\it Spitzer}-IRS spectra of
{\it typical} protoplanetary disks around solar-mass stars (from top to bottom, AA Tau, AS 205 and DR Tau, scaled for clarity, from \cite{Carr08,Salyk08}). 
Note that the model only includes water lines, not the Q branches from C$_2$H$_2$, HCN and CO$_2$, which
are prominent in the data. The observed spectra show
lines of many more molecular and atomic species.  }
\label{RADLITE_fullrange}
\end{figure}

\subsection{The CO fundamental}
\label{COvib}

Given the large volume of observations of the CO fundamental, it is expected that
modeling of these lines will be a central application of RADLite. CO ro-vibrational lines are ubiquitous tracers of
disk surfaces at $\sim 1\,$AU \citep{Najita03,Blake04,Brittain07,Salyk07}. Some disks are already being
resolved in CO ro-vibrational lines with AO-fed spectrometers on 8-10\,m class telescopes \citep{Goto06}. With the improvement
in spatial resolution of an ELT by factors of 3-5, the CO emission from many disks should be imaged 
over a correspondingly larger number of spaxels. 
Further, the CO fundamental can be used to demonstrate the ability of the code to simultaneously treat
dynamic ranges of 4-5 orders of magnitude in both velocity and spatial dimension. Figure \ref{co_inclined} shows a rendering of the $^{12}$CO
band (lines from R(0)-P(14)) for the fiducial disk model viewed at high inclination ($\sim 70\degr$). 
The models exhibit double-peaked emission line profiles typical for Keplerian disks. The lines are 
formed in the inner 1 AU of the disk. Superposed on the emission lines are
very narrow absorption lines formed as continuum and line emission from the inner disk is absorbed by the outer disk -- a configuration made
possible by a flaring disk geometry. Such inclined disks have been used in the past to study disk material
through absorption features \citep{Pontoppidan05,Brittain05,Lahuis06,Gibb07}. 
Two different models are produced, for inclinations of 68 and 72 degrees, illustrating 
how rapidly the CO absorption changes with inclination; essentially no absorption is seen at inclinations below 68 degrees, while
the central continuum becomes very faint ($\ll 0.1\,$Jy) at inclinations higher than 75 degrees. The prediction is that
only $\sim$5\% of disks will have the appropriate inclination to show absorption from the outer disk. 
Qualitatively, the model spectrum at 72 degrees inclination resembles that of HL Tau, a disk suggested to have a 
large gas-to-dust ratio in its upper layers, possibly to due to dust settling \citep{Brittain05}.

There are currently no high resolution IFU spectrometers that operate in the 3-20\,$\mu$m region. However, 
one imaging technique that has been used for CO is spectro-astrometry \citep{Pontoppidan08,vanderplas09}. 
Figure \ref{co_sa} shows astrometric spectra at three different position angles (PAs) 
of the $i=72\degr$ P(8) CO ro-vibrational line also seen in Figure \ref{co_inclined}. 
At a PA=0$\degr$, the classical antisymmetric profile of a Keplerian disk appears. The absorption from the 
outer disk shows up in the astrometry as a second antisymmetric profile -- but with the opposite sign of that produced by the emission. 
This signature is due to the Keplerian rotation of the outer disk having a significantly varying radial component across the 
emission source (which consists of the central star, thermal emission from the innermost part of the disk, as well as scattering 
from the disk surface), and can therefore potentially be used to distinguish CO absorption from the disk, as opposed to
CO absorption from an unrelated foreground cloud (which would have no astrometric signature). 
Also notable is a significant asymmetry seen in the off-axis angles 60 and 120$\degr$. For instance, at 60$\degr$, the 
blue-shifted lobe has a much smaller offset than the red-shifted lobe. This asymmetry is due
to a combination of two effects. First, the structure of the inner rim in the disk causes an asymmetry
of the continuum. When viewed at an inclined angle, the 
photocenter of the continuum is shifted in the direction of the far side of the disk. Because spectro-astrometry is
a relative measurement of a line compared to a continuum, this effect takes the appearance of a line asymmetry. Second, there
is a true line asymmetry due to the fact that the disk is not perfectly flat. Thus, the surface isovelocity contours are
not symmetric around the disk major axis and the angle of a ray with respect to the disk surface normal
is larger from the near side of the disk relative to the far side. 
The asymmetries are strong enough to be easily detectable with current spectro-astrometric facilities such as CRIRES
on the Very Large Telescope (VLT), and can be effectively modeled
with RADLite.

The large apertures of the ELTs will be able to produce line images of planet-forming regions across many 
spaxels. The fundamental ro-vibrational band of CO will be the easiest 
imaging target. The expected improvement of a 42\,m telescope relative to an 8.2\,m telescope 
is illustrated in Figs. \ref{SR21} and \ref{ELT_CO}. The former shows an actual single-slit observation of a CO line
from the protoplanetary disk SR 21, which has an evolved planet-forming region characterized by a large inner hole (a transitional or cold disk) 
\citep[see][for details]{Pontoppidan08}. The observation
is modeled with RADLite using the basic disk geometry derived by \cite{Pontoppidan08}, who determined that
the CO gas is truncated within 6.5\,AU. The additional applicability of RADLite is illustrated by simulating observations using two 
different gas-to-dust ratios to simulate the effect of grain growth. The low gas-to-dust ratio is 1280 while the
high gas-to-dust ratio of $2.56\times 10^5$ has lowered the disk opacity sufficiently to make the disk optically thin to the continuum
emission at 4.7\,$\mu$m. This produces two emission lobes in the position-velocity image, in contrast to the optically
thick disk, which results in a single elongated structure. This difference is due to a limb brightening effect of the
inner rim of the disk. For an optically thick inner rim, the surface intensity of the inner wall is constant at different ray impact parameters;
the inner rim acts as an opaque ``wall''. An optically thin disk will result in a higher intensity at image locations where a ray is 
near tangential to the inner rim, i.e. where a ray samples the largest column density of warm gas. While the observation
hints at an optically thin disk, the greater sensitivity and higher spatial resolution of an ELT observation will
enable such conclusions to be made with much greater confidence and for a much wider range of disks. 

It is likely that an ELT mid-infrared instrument will not be a single-slit facility, but an IFU.
Figure \ref{ELT_CO} shows a RADLite simulation
of the same disk as observed with a 42\,m aperture telescope equipped with an IFU.
In this case, the CO emission would be imaged across at least 50 spaxels within 12\,AU. It will thus be possible to search for
kinematic structures in the disk directly related to planet formation, such as spiral waves and the radial density structure in
the optically thin parts of the disk.

\begin{figure*}
\centering
\includegraphics[width=16cm,angle=0]{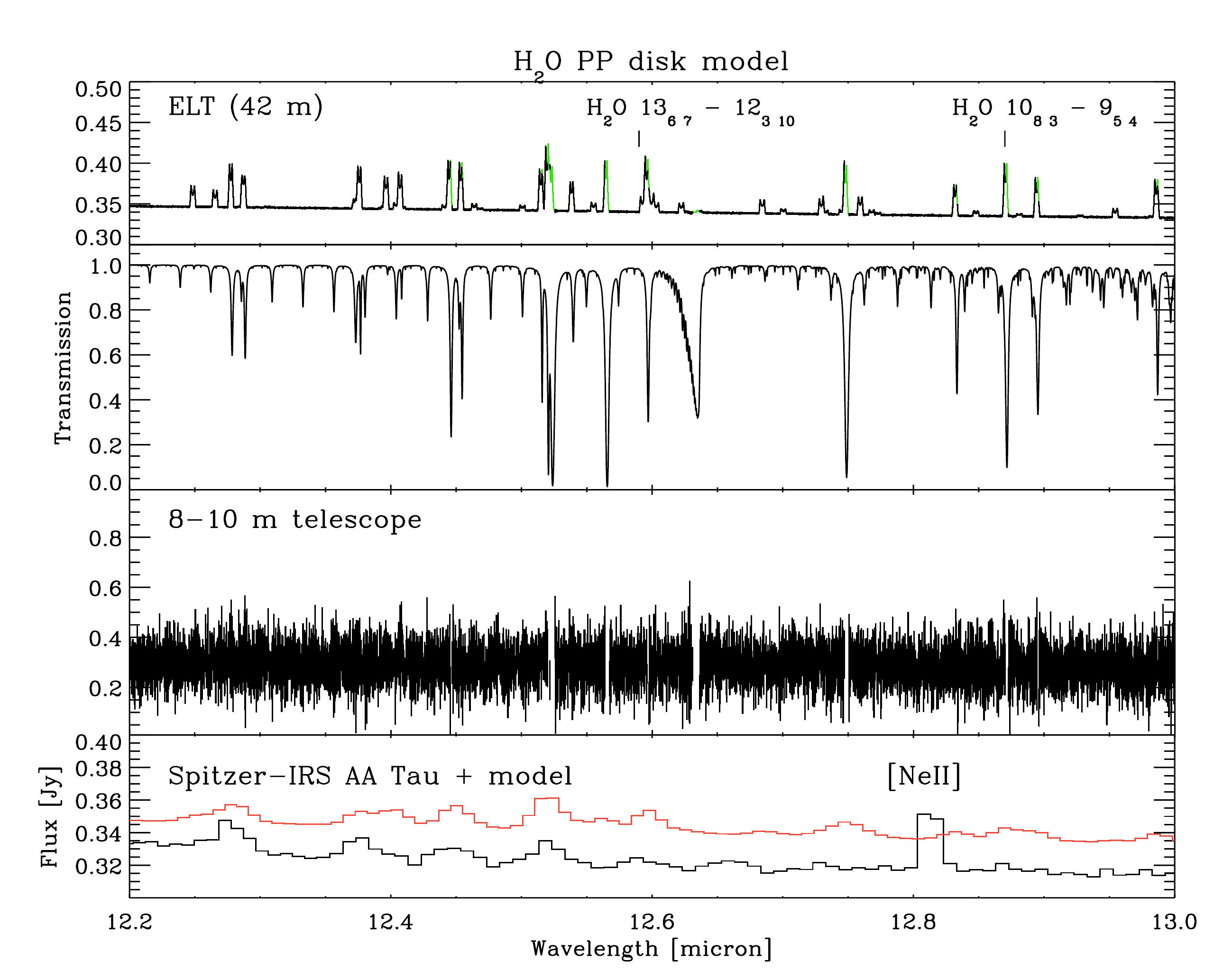}
\caption[]{Simulations of ground-based observations of water lines from a typical protoplanetary disk. Top panel:
model spectrum observed with a resolving power of $R=\lambda/\Delta\lambda = 50,000$ with a conservative noise estimate for the E-ELT (3 times
the theoretical estimate). Numerous strong lines (10-20\% relative to the continuum in this particular model) 
are easily measured through the atmospheric transmission (second panel from the top). The black curve is the model spectrum with saturated ($<$50\% transmission)
parts of the telluric spectrum cut out. Third panel from the top: The same model spectrum, but now for the sensitivity ($\sigma=$100 mJy)
of VLT-VISIR at $\lambda/\Delta\lambda = 30,000$. Bottom panel: The observed {\it Spitzer} spectrum of AA Tau (lower black curve) relative to the
same RADLite model (upper red curve) convolved to the resolution of the short-high module of {\it Spitzer}-IRS ($\lambda/\Delta\lambda = 600$). 
All sensitivities assume roughly 1 hour of on-source integration time. Note that AA Tau is not a particularly bright
target; with a continuum flux of 0.3\,Jy at 10\,$\mu$m, hundreds of protoplanetary disks will be brighter, some by
an order of magnitude.}
\label{ELT_SPEC}
\end{figure*}

\begin{figure}
\centering
\includegraphics[width=9cm]{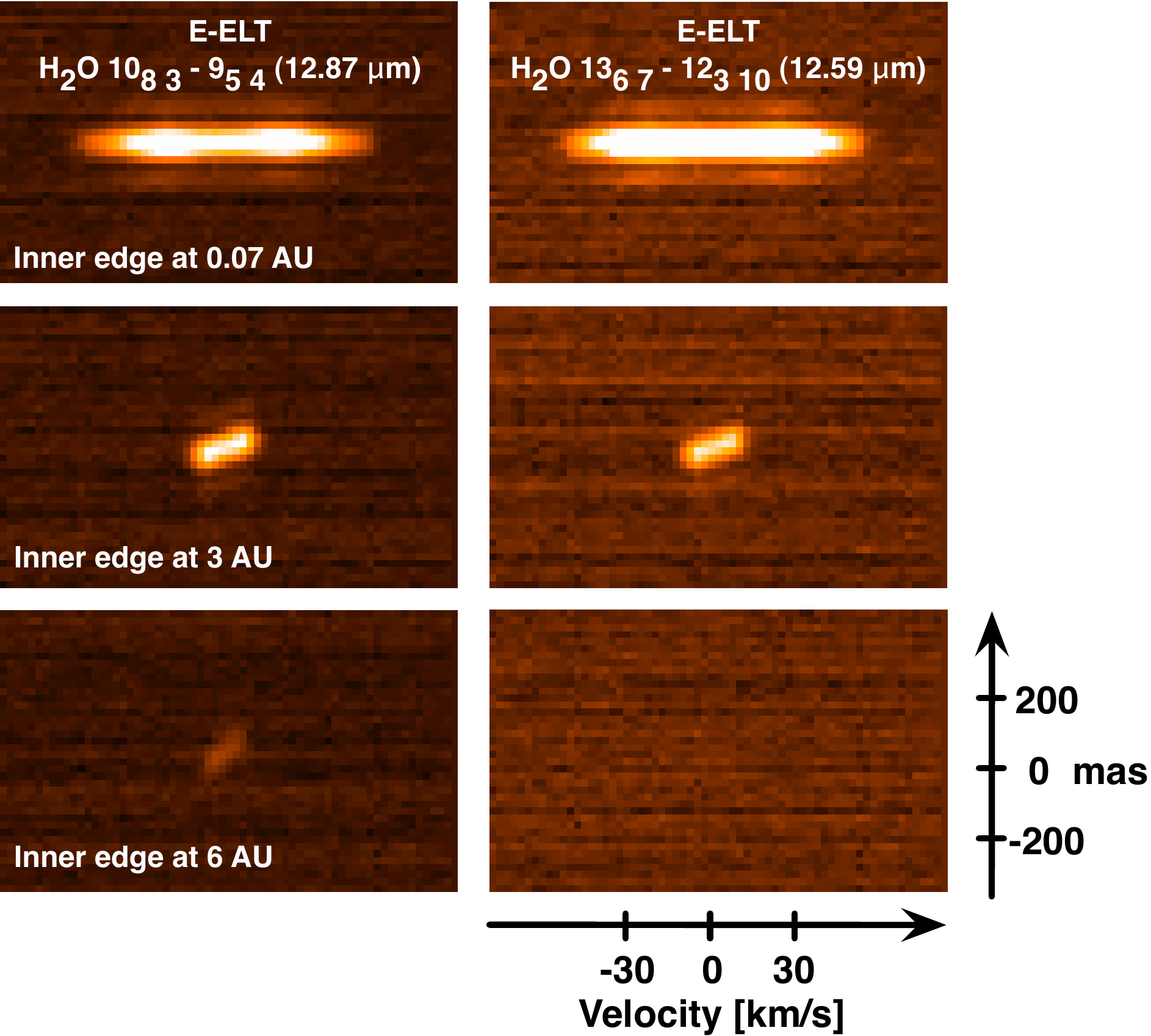}
\caption[]{Simulated E-ELT continuum-subtracted echellograms of two N-band H$_2$O lines for the fiducial non-LTE H$_2$O model, 
both of which require velocity shifts of $\sim 30\,\rm km\,s^{-1}$ to shift the lines out of
the telluric absorption line. The left line has an upper level energy of 3243\,K, the 
right has an upper level energy of 3966\,K. In the lower two rows of panels, models with
a 3 and 6\,AU inner hole are also calculated. For clarity, the atmospheric transmission is not simulated. }
\label{H2O_imaging}
\end{figure}

\begin{table*}
\centering
\caption{Strong N-band H$_2$O lines suitable for ground-based observations}
\begin{tabular}{lllllll}
\hline
\hline
Transition\tablenotemark{a}   & & Wavelength [$\mu$m]\tablenotemark{b} & Frequency [cm$^{-1}$] & {\it Spitzer} detected\tablenotemark{c} & Strength\tablenotemark{d} & Transmission\tablenotemark{e}  \\
\hline

o $v=1-0\ \ 17_{0\ 17}$ & $\rightarrow\ 18_{1\ 18}$  & 8.09583  & 1235.20415 & N/A & 0.09 & clean \\
o $v=1-0\ \ 11_{8\ 3}$  & $\rightarrow\ 12_{9\ 4}$   & 8.10842  & 1233.28658 & N/A & 0.09 & clean \\
o $v=1-0\ \ 12_{9\ 4}$  & $\rightarrow\ 13_{10\ 3}$  & 8.19373  & 1220.44594 & N/A & 0.09 & clean \\
o $v=1-0\ \ 11_{6\ 5}$  & $\rightarrow\ 12_{7\ 6}$   & 8.19441  & 1220.34459 & N/A & 0.05 & clean \\
p $v=1-0\ \ 11_{6\ 6}$  & $\rightarrow\ 12_{7\ 5}$   & 8.20245  & 1219.14720 & N/A & 0.05 & clean \\
p $v=1-0\ \ 18_{0\ 18}$ & $\rightarrow\ 19_{1\ 19}$  & 8.23469  & 1214.37420 & N/A & 0.08 & clean \\
p $v=1-0\ \ 11_{5\ 7}$  & $\rightarrow\ 12_{6\ 6}$   & 8.24397  & 1213.00716 & N/A & 0.07 & clean \\
o $v=1-0\ \ 12_{8\ 5}$  & $\rightarrow\ 13_{9\ 4}$   & 8.26608  & 1209.76296 & N/A & 0.06 & clean \\
o $v=1-0\ \ 12_{7\ 6}$  & $\rightarrow\ 13_{8\ 5}$   & 8.32319  & 1201.46258 & N/A & 0.05 & clean  \\
o $v=0-0\ \ 13_{6\ 7}$  & $\rightarrow\ 12_{1\ 12}$  & 8.34329  & 1198.56748 & N/A & 0.05 & clean \\
o $v=1-0\ \ 12_{6\ 7}$  & $\rightarrow\ 13_{7\ 6}$   & 8.36579  & 1195.34395 & N/A & 0.06 & clean \\
o $v=1-0\ \ 13_{8\ 5}$  & $\rightarrow\ 14_{9\ 6}$   & 8.42794  & 1186.52920 & N/A & 0.05 & clean \\
o $v=1-0\ \ 12_{5\ 8}$  & $\rightarrow\ 13_{6\ 7}$   & 8.44520  & 1184.10422 & N/A & 0.06 & clean \\
o $v=1-0\ \ 12_{4\ 9}$  & $\rightarrow\ 13_{5\ 8}$   & 8.76006  & 1141.54447 & N/A & 0.05 & clean \\
o $v=1-0\ \ 9_{2\ 7}$   & $\rightarrow\ 10_{5\ 6}$   & 9.09356  & 1099.67926 & N/A & 0.06 & clean \\
o $v=1-0\ \ 10_{2\ 9}$  & $\rightarrow\ 11_{3\ 8}$   & 9.16418  & 1091.20494 & N/A & 0.06 & clean \\
o $v=0-0\ \ 14_{6\ 9}$  & $\rightarrow\ 13_{1\ 2}$   & 9.59209  & 1042.52515 & N/A & 0.05 & clean \\
o $v=0-0\ \ 19_{2\ 17}$ & $\rightarrow\ 18_{1\ 18}$  & 9.88515  & 1011.61860 & N/A & 0.09 & clean \\
o $v=0-0\ \ 14_{10\ 5}$ & $\rightarrow\ 13_{7\ 6}$   & 10.05642 & 994.38945  & No  & 0.05 & clean \\
o $v=0-0\ \ 17_{7\ 10}$ & $\rightarrow\ 16_{4\ 13}$  & 10.11319 & 988.80730  & No  & 0.05 & clean \\
o $v=0-0\ \ 17_{8\ 9}$  & $\rightarrow\ 16_{5\ 12}$  & 10.27482 & 973.25273  & No  & 0.05 & clean \\
o $v=0-0\ \ 13_{10\ 3}$ & $\rightarrow\ 12_{7\ 6}$   & 10.30317 & 970.57472  & No  & 0.05 & clean \\
o $v=0-0\ \ 18_{3\ 16}$ & $\rightarrow\ 17_{1\ 17}$  & 10.42505 & 959.22783  & No  & 0.10 & clean \\
o $v=0-0\ \ 16_{9\ 8}$  & $\rightarrow\ 15_{6\ 9}$   & 10.47955 & 954.23969  & No  & 0.05 & clean \\
o $v=0-0\ \ 12_{5\ 8}$  & $\rightarrow\ 11_{0\ 11}$  & 10.54560 & 948.26294  & No  & 0.06 & clean \\
o $v=0-0\ \ 14_{9\ 6}$  & $\rightarrow\ 13_{6\ 7}$   & 10.76434 & 928.99316  & No  & 0.07 & clean \\
o $v=0-0\ \ 12_{7\ 6}$  & $\rightarrow\ 11_{2\ 9}$   & 10.84439 & 922.13542  & No  & 0.08 & clean \\
o $v=0-0\ \ 15_{6\ 9}$  & $\rightarrow\ 14_{3\ 12}$  & 10.85308 & 921.39780  & No  & 0.09 & clean \\
o $v=0-0\ \ 13_{9\ 4}$  & $\rightarrow\ 12_{6\ 7}$   & 10.94112 & 913.98290  & No  & 0.08 & clean \\
o $v=0-0\ \ 19_{4\ 15}$ & $\rightarrow\ 18_{3\ 16}$  & 10.97414 & 911.23324  & No  & 0.07 & clean \\
o $v=0-0\ \ 18_{4\ 15}$ & $\rightarrow\ 17_{1\ 16} $ & 10.98045 & 910.70960  & No  & 0.09 & clean   \\
o $v=0-0\ \ 12_{6\ 7} $ & $\rightarrow\ 11_{1\ 10} $ & 11.00168 & 908.95248  & No  & 0.10 & $>$30 $\rm km\,s^{-1}$  \\
o $v=0-0\ \ 15_{8\ 7} $ & $\rightarrow\ 14_{5\ 10} $ & 11.02836 & 906.75277  & No  & 0.10 & $>$30 $\rm km\,s^{-1}$  \\
p $v=0-0\ \ 17_{3\ 15}$ & $\rightarrow\ 16_{0\ 16} $ & 11.03376 & 906.30900  & No  & 0.05 & clean   \\
o $v=0-0\ \ 17_{2\ 15}$ & $\rightarrow\ 16_{1\ 16} $ & 11.03475 & 906.22805  & No  & 0.10 & clean   \\
p $v=0-0\ \ 12_{9\ 3}$  & $\rightarrow\ 11_{6\ 6}$   & 11.23479 & 890.09255  & No  & 0.05 & clean \\
o $v=0-0\ \ 12_{9\ 4}$  & $\rightarrow\ 11_{6\ 5}$   & 11.25311 & 888.64354  & No  & 0.10 & clean \\
o $v=0-0\ \ 15_{7\ 8} $ & $\rightarrow\ 14_{4\ 11} $ & 11.32409 & 883.07302  & No  & 0.10 & clean   \\
p $v=0-0\ \ 11_{5\ 7} $ & $\rightarrow\ 10_{0\ 10} $ & 11.47773 & 871.25270  & No  & 0.05 &  clean   \\
o $v=0-0\ \ 11_{9\ 2}$  & $\rightarrow\ 10_{6\ 5}  $ & 11.55471 & 865.44774  & No  & 0.10 & $>$30 $\rm km\,s^{-1}$  \\
p $v=0-0\ \ 11_{9\ 3}$  & $\rightarrow\ 10_{6\ 4}  $ & 11.56124 & 864.95886  & No  & 0.05 & $>$30 $\rm km\,s^{-1}$  \\
o $v=0-0\ \ 17_{3\ 14}$ & $\rightarrow\ 16_{2\ 15}  $& 11.64764 & 858.54294  & No  & 0.12 &  clean  \\
p $v=0-0\ \ 14_{6\ 8} $ & $\rightarrow\ 14_{6\ 8} $  & 11.71816 & 853.37648  & No  & 0.08 &  clean   \\
o $v=0-0\ \ 16_{3\ 14}$ & $\rightarrow\ 15_{0\ 15} $ & 11.72455 & 852.91119  & Yes & 0.15 &  $>$30 $\rm km\,s^{-1}$  \\
o $v=0-0\ \ 16_{8\ 9} $ & $\rightarrow\ 15_{5\ 10} $ & 11.82409 & 845.73138  & No  & 0.07 & clean   \\
p $v=0-0\ \ 15_{8\ 8} $ & $\rightarrow\ 14_{5\ 9} $  & 11.88916 & 841.10204  & No  & 0.06 & clean   \\
p $v=0-0\ \ 14_{7\ 7} $ & $\rightarrow\ 13_{4\ 10} $ & 11.90026 & 840.31789  & No  & 0.09 & clean   \\
o $v=0-0\ \ 14_{8\ 7} $ & $\rightarrow\ 13_{5\ 8}  $ & 11.96812 & 835.55295  & No  & 0.10 & clean   \\
p $v=0-0\ \ 13_{8\ 6} $ & $\rightarrow\ 12_{5\ 7}  $ & 12.09017 & 827.11790  & No  & 0.08 & clean   \\
p $v=0-0\ \ 17_{5\ 13}$ & $\rightarrow\ 16_{2\ 14} $ & 12.24814 & 816.45022  & Yes & 0.07 & clean   \\
o $v=0-0\ \ 12_{8\ 5} $ & $\rightarrow\ 11_{5\ 6}  $ & 12.27721 & 814.51703  & Yes & 0.15 & $>$30 $\rm km\,s^{-1}$  \\
o $v=0-0\ \ 16_{4\ 13}$ & $\rightarrow\ 15_{1\ 14} $ & 12.37566 & 808.03788  & Yes & 0.15 & $>$30 $\rm km\,s^{-1}$  \\
o $v=0-0\ \ 17_{4\ 13}$ & $\rightarrow\ 16_{3\ 14} $ & 12.39625 & 806.69564  & Yes & 0.12 & clean   \\
p $v=0-0\ \ 16_{3\ 13}$ & $\rightarrow\ 15_{2\ 14} $ & 12.40705 & 805.99363  & Yes & 0.13 & clean   \\
o $v=0-0\ \ 11_{8\ 3} $ & $\rightarrow\ 10_{5\ 6}  $ & 12.44483 & 803.54638  & Yes & 0.17 & $>$30 $\rm km\,s^{-1}$  \\
o $v=0-0\ \ 13_{7\ 6} $ & $\rightarrow\ 12_{4\ 9}  $ & 12.45346 & 802.98996  & Yes & 0.17 & $>$30 $\rm km\,s^{-1}$  \\
p $v=0-0\ \ 15_{3\ 13}$ & $\rightarrow\ 14_{0\ 14} $ & 12.51462 & 799.06551  & Yes & 0.15 & $>$30 $\rm km\,s^{-1}$\\
p $v=0-0\ \ 11_{8\ 4} $ & $\rightarrow\ 10_{5\ 5} $  & 12.53831 & 797.55588  & Yes & 0.10 & clean\\
o $v=0-0\ \ 13_{6\ 7} $ & $\rightarrow\ 12_{3\ 10} $ & 12.59591 & 793.90838  & Yes & 0.20 & $>$30 $\rm km\,s^{-1}$  \\
p $v=0-0\ \ 10_{8\ 2} $ & $\rightarrow\ 9_{5\ 5} $   & 12.83197 & 779.30366  & No  & 0.12 & $>$30 $\rm km\,s^{-1}$  \\
p $v=0-0\ \ 12_{5\ 7} $ & $\rightarrow\ 11_{2\ 10} $ & 12.89409 & 775.54902  & No  & 0.15 & $>$30 $\rm km\,s^{-1}$  \\
p $v=0-0\ \ 12_{7\ 5} $ & $\rightarrow\ 11_{4\ 8}  $ & 12.98575 & 770.07497  & No  & 0.15 & $>$30 $\rm km\,s^{-1}$  \\
o $v=0-0\ \ 16_{5\ 12}$ & $\rightarrow\ 15_{2\ 13} $ & 13.03333 & 767.26380  & Yes & 0.12 & clean   \\
o $v=0-0\ \ 16_{7\ 10}$ & $\rightarrow\ 15_{4\ 11} $ & 13.13245 & 761.47255  & Yes & 0.09 & clean   \\
o $v=0-0\ \ 15_{3\ 12}$ & $\rightarrow\ 14_{2\ 13} $ & 13.29319 & 752.26468  & Yes & 0.20 & $>$30 $\rm km\,s^{-1}$  \\
\hline
\end{tabular}
\tablenotetext{a}{Bending mode vibrational quantum number.}
\tablenotetext{b}{Line parameters from HITRAN2004 \citep{Rothman05}.}
\tablenotetext{c}{Subjective estimate limited by severe line blending.}
\tablenotetext{d}{Peak line flux/continuum flux.}
\tablenotetext{e}{Is a significant Doppler shift required to observe the transition?}

\label{line_table}
\end{table*}

\subsection{H$_2$O lines in the atmospheric $N$ band window}
\label{H2O}
An important new window for high resolution
spectroscopy that will be opened by ELT class telescopes is the 7.5-13.9\,$\mu$m atmospheric N-band, as well as the 17-25\,$\mu$m atmospheric Q-band.
The only reason that these windows are currently not very productive for ground-based high resolution spectroscopic 
work, is that the sensitivity of instruments on 8-10\,m class telescopes is not sufficient to reach the vast majority of protoplanetary disks, in 
particular disks around T Tauri stars now known to exhibit the richest molecular emission spectra.
(See also recent work by \cite{Knez09} and \cite{Najita09} for discussions of the power and limitations of current high resolution N- and Q-band spectroscopy). 
The massive improvements in sensitivity offered by an ELT will put hundreds or thousands of protoplanetary disks within reach for high resolution imaging
spectroscopy in the N- and Q-bands.  

There are many atomic and molecular species with strong transitions in the N- and Q-bands. One that has received considerable recent 
attention is the [NeII] 12.8\,$\mu$m line \citep{Pascucci07,vanBoekel09,Najita09}. 
Here, we show that a large number of rotational lines of water can be reached from the ground. These
are lines that are known to be strong in many protoplanetary disks \citep[][and work in preparation by these authors]{Carr08,Salyk08}. 

The strong lines and the associated high apparent molecular abundances
in the disk surfaces within a few AU were not anticipated by the initial chemical
models -- H$_2$O abundances in the warm surface layers appear
to be of order $n_{\rm H_2O} / n_{\rm H}\sim 10^{-5} - 10^{-4}$, while
many contemporary models \citep{Glassgold04, Gorti08} predicted
significantly less water -- $n_{\rm H_2O} / n_{\rm H}\sim 10^{-8} -
10^{-6}$. Other modelers are now working on explaining very high molecular abundances, including water
\citep{Glassgold09,Woods09}. 

Strong lines can be found across the entire 10-36\,$\mu$m {\it Spitzer} high resolution spectral range 
and, by extension, many strong lines are expected in the 6\,$\mu$m water bending mode as well as in the 40-200\,$\mu$m range, 
in the regime of future space-borne observatories. {\it Herschel}-PACS
will cover lines between 60 and 210\,$\mu$m, but will not spectrally resolve them. 
The first facilities with the potential to produce spatially and spectrally resolved data of rotational water lines from a significant sample
of disks will be the ELTs. Figure \ref{RADLITE_fullrange} shows an overview of 
water lines along with near-term and future infrared facilities. The most sensitive ground-based window for observing the
lines is the atmospheric $N$-band window, which extends from 7.5\,$\mu$m to 13.9\,$\mu$m.
Figure \ref{ELT_SPEC} shows a RADLite model spectrum of the fiducial model, {\it based on
observed Spitzer spectra}, of part of the $N$ band. The model includes atmospheric absorption for 
a dry site at 2600\,m elevation. It also assumes that regions of the spectrum with less than 
50\% transmission cannot be calibrated, and these parts of the spectrum have been removed. As with
most molecular infrared spectroscopy from the ground, it is necessary to observe the target at a time
when the Doppler shift due to the relative motion of the Earth is large in order to
move the target line away from its telluric counterpart. In this instance, a typical shift
of 30\,$\rm km\,s^{-1}$ has been assumed. Finally, noise relevant for current 
mid-infrared high resolution spectrometers as well as for an instrument mounted on 
an ELT has been added to the spectra. It can be seen that for this particular model, approximately 
50 H$_2$O lines will be observable from the ground under dry conditions (1-3 mm of precipitable water vapor). Most line profiles
can be measured in their entirety in one observing epoch, while a few of the lower excitation lines will require the combination
of two epochs with opposite Doppler velocity shifts, i.e. separated by 6 months. The improvement in sensitivity of 
an ELT as compared to existing ground-based facilities clearly moves this type of observation from being very difficult to 
being straightforward for a large sample of disks, given that the model example is of typical brightness for nearby T Tauri stars (0.35\,Jy at 10\,$\mu$m). 

Figure \ref{H2O_imaging} shows simulations of H$_2$O line imaging with the E-ELT, assuming a single-slit instrument. The figure
shows three different versions of the fiducial non-LTE model, the first with the inner disk truncated at the dust sublimation radius of 0.07\,AU, the
other two assuming inner holes with radii of 3 and 6\,AU. The echellograms show that a 3\,AU hole is easily spatially resolved. Two rotational lines
are modeled: $10_{8\ 3}\rightarrow 9_{5\ 4}$ and $13_{6\ 7}\rightarrow 12_{3\ 10}$, with upper level excitation
energies of 3243\,K and 3966\,K, respectively (for comparison, the excitation energy of the CO ro-vibrational fundamental is $\sim 3100$\,K). The low
excitation line can be detected out to 6 AU, while the higher excitation line drops below
the detection limit between 3 and 6 AU. Astrometric spectra also detect the offsets for the 0.07\,AU truncation radius. 
This shows that some lines are better tracers of certain radii than others due to a balance between excitation temperatures and geometric enhancement, i.e.
the higher excitation line is the brightest at small radii but the faintest at large radii. 

In Table \ref{line_table}, an incomplete list of molecular lines most suitable for 
observations from the ground is given. The list is compiled based on 
line brightness (more than 5\% relative to the continuum) and atmospheric transparency (more than 50\% transmission
over at least half the line, assuming 30\,$\rm km\,s^{-1}$ velocity shift relative to the atmosphere). 
Given that this selection is based on a model of one specific source, in practice, potentially thousands of molecular lines in 3-28\,$\mu$m range will
be detectable from the ground with an ELT equipped with a high resolution spectrometer. 

\section{Conclusions}
\label{conclusion}
We have presented a new axisymmetric raytracer, called RADLite, for simulating telescopic line images, with specific emphasis on
infrared lines formed in the planet-forming region of protoplanetary disks, and applied it to mid-infrared 
CO and H$_2$O emission. RADLite is optimized for
rapidly rendering large numbers of lines for full 2D axisymmetric structures with arbitrarily large velocity gradients, including
a detailed treatment of the dust radiative transfer. For instance, a full spectrum of $\sim$1000 water lines in the infrared (2-200\,$\mu$m), can
be rendered with a velocity resolution of 1~$\rm km\,s^{-1}$ in 1-2 hours on a single workstation. The code has very general applications to
chemical and excitation disk models as well as observations from infrared spectrometers on 8-10\,m class ground-based telescopes, {\it Spitzer}-IRS, 
{\it Herschel}-PACS, future facilities such as the JWST, SOFIA, E-ELT, TMT and the Giant Magellan Telescope (GMT), in addition to
(sub)millimeter facilities, such as ALMA. While not explicitly modeled
in this paper, we also expect applications to future infrared interferometer facilities, such as VLT-MATISSE. We find that a
primary reason that infrared molecular spectroscopy of disks in the N-band (H$_2$O and a wide range of other species) has not received much
attention so far is due to a sensitivity deficit of roughly a factor of 10. This will be remedied by the ELT generation and the JWST, 
and we predict that this will become a highly active research area in the future. 
The potential applications for RADLite reach beyond protoplanetary disks. Any axisymmetric structure can be modeled. This
could potentially include protoplanetary nebulae, the atmospheres of asymptotic giant branch stars, and active galactic nuclei.

\acknowledgments{An anonymous referee provided a thorough and constructive report that improved
the manuscript considerably. The authors are grateful to Sarah Kendrew, Leiden Observatory, for
providing a high resolution infrared transmission spectrum model. Discussions with
Joan Najita, Colette Salyk and Marco Spaans have been valuable during the preparation of the manuscript.
Support for KMP was provided by NASA through Hubble Fellowship grant \#01201.01 
awarded by the Space Telescope Science Institute, which is operated by the Association of 
Universities for Research in Astronomy, Inc., for NASA, under contract NAS 5-26555. RM has been
supported by NSF grant AST-0708922. 
}

\bibliographystyle{apj}
\bibliography{ms}

\begin{thebibliography}{66}
\expandafter\ifx\csname natexlab\endcsname\relax\def\natexlab#1{#1}\fi

\bibitem[{{Acke} {et~al.}(2005){Acke}, {van den Ancker}, \&
  {Dullemond}}]{Acke05}
{Acke}, B., {van den Ancker}, M.~E., \& {Dullemond}, C.~P. 2005, \aap, 436, 209

\bibitem[{{Ayliffe} \& {Bate}(2009)}]{Ayliffe09}
{Ayliffe}, B.~A., \& {Bate}, M.~R. 2009, \mnras, 393, 49

\bibitem[{{Blake} \& {Boogert}(2004)}]{Blake04}
{Blake}, G.~A., \& {Boogert}, A.~C.~A. 2004, \apjl, 606, L73

\bibitem[{{Bockelee-Morvan} {et~al.}(1998){Bockelee-Morvan}, {Gautier}, {Lis},
  {Young}, {Keene}, {Phillips}, {Owen}, {Crovisier}, {Goldsmith}, {Bergin},
  {Despois}, \& {Wootten}}]{Bockelee-Morvan98}
{Bockelee-Morvan}, D. {et~al.} 1998, Icarus, 133, 147

\bibitem[{{Brittain} {et~al.}(2005){Brittain}, {Rettig}, {Simon}, \&
  {Kulesa}}]{Brittain05}
{Brittain}, S.~D., {Rettig}, T.~W., {Simon}, T., \& {Kulesa}, C. 2005, \apj,
  626, 283

\bibitem[{{Brittain} {et~al.}(2007){Brittain}, {Simon}, {Najita}, \&
  {Rettig}}]{Brittain07}
{Brittain}, S.~D., {Simon}, T., {Najita}, J.~R., \& {Rettig}, T.~W. 2007, \apj,
  659, 685

\bibitem[{{Brown} {et~al.}(2007){Brown}, {Blake}, {Dullemond}, {Mer{\'{\i}}n},
  {Augereau}, {Boogert}, {Evans}, {Geers}, {Lahuis}, {Kessler-Silacci},
  {Pontoppidan}, \& {van Dishoeck}}]{Brown07}
{Brown}, J.~M. {et~al.} 2007, \apjl, 664, L107

\bibitem[{{Calvet} {et~al.}(2002){Calvet}, {D'Alessio}, {Hartmann}, {Wilner},
  {Walsh}, \& {Sitko}}]{Calvet02}
{Calvet}, N., {D'Alessio}, P., {Hartmann}, L., {Wilner}, D., {Walsh}, A., \&
  {Sitko}, M. 2002, \apj, 568, 1008

\bibitem[{{Canup} \& {Ward}(2002)}]{Canup02}
{Canup}, R.~M., \& {Ward}, W.~R. 2002, \aj, 124, 3404

\bibitem[{{Carr} \& {Najita}(2008)}]{Carr08}
{Carr}, J.~S., \& {Najita}, J.~R. 2008, Science, 319, 1504

\bibitem[{{Chabrier} {et~al.}(2007){Chabrier}, {Baraffe}, {Selsis}, {Barman},
  {Hennebelle}, \& {Alibert}}]{Chabrier07}
{Chabrier}, G., {Baraffe}, I., {Selsis}, F., {Barman}, T.~S., {Hennebelle}, P.,
  \& {Alibert}, Y. 2007, in Protostars and Planets V, ed. B.~{Reipurth},
  D.~{Jewitt}, \& K.~{Keil}, 623--638

\bibitem[{{Chiang} \& {Goldreich}(1997)}]{CG97}
{Chiang}, E.~I., \& {Goldreich}, P. 1997, \apj, 490, 368

\bibitem[{{Dullemond} \& {Dominik}(2004)}]{Dullemond04}
{Dullemond}, C.~P., \& {Dominik}, C. 2004, \aap, 417, 159

\bibitem[{{Dullemond} \& {Turolla}(2000)}]{Dullemond00}
{Dullemond}, C.~P., \& {Turolla}, R. 2000, \aap, 360, 1187

\bibitem[{{Encrenaz}(2008)}]{Encrenaz08}
{Encrenaz}, T. 2008, \araa, 46, 57

\bibitem[{{Evans} {et~al.}(2008){Evans}, {Dunham}, {J{\o}rgensen}, {Enoch},
  {Mer{\'{\i}}n}, {van Dishoeck}, {Alcal{\'a}}, {Myers}, {Stapelfeldt},
  {Huard}, {Allen}, {Harvey}, {van Kempen}, {Blake}, {Koerner}, {Mundy},
  {Padgett}, \& {Sargent}}]{Evans09}
{Evans}, II, N.~J. {et~al.} 2008, ArXiv e-prints

\bibitem[{{Fedele} {et~al.}(2008){Fedele}, {van den Ancker}, {Acke}, {van der
  Plas}, {van Boekel}, {Wittkowski}, {Henning}, {Bouwman}, {Meeus}, \&
  {Rafanelli}}]{Fedele08}
{Fedele}, D. {et~al.} 2008, \aap, 491, 809

\bibitem[{{Gibb} {et~al.}(2007){Gibb}, {Van Brunt}, {Brittain}, \&
  {Rettig}}]{Gibb07}
{Gibb}, E.~L., {Van Brunt}, K.~A., {Brittain}, S.~D., \& {Rettig}, T.~W. 2007,
  \apj, 660, 1572

\bibitem[{{Glassgold} {et~al.}(2009){Glassgold}, {Meijerink}, \&
  {Najita}}]{Glassgold09}
{Glassgold}, A.~E., {Meijerink}, R., \& {Najita}, J.~R. 2009, ArXiv e-prints

\bibitem[{{Glassgold} {et~al.}(2004){Glassgold}, {Najita}, \&
  {Igea}}]{Glassgold04}
{Glassgold}, A.~E., {Najita}, J., \& {Igea}, J. 2004, \apj, 615, 972

\bibitem[{{Gorti} \& {Hollenbach}(2008)}]{Gorti08}
{Gorti}, U., \& {Hollenbach}, D. 2008, \apj, 683, 287

\bibitem[{{Goto} {et~al.}(2006){Goto}, {Usuda}, {Dullemond}, {Henning}, {Linz},
  {Stecklum}, \& {Suto}}]{Goto06}
{Goto}, M., {Usuda}, T., {Dullemond}, C.~P., {Henning}, T., {Linz}, H.,
  {Stecklum}, B., \& {Suto}, H. 2006, \apj, 652, 758

\bibitem[{{Hartmann} {et~al.}(1998){Hartmann}, {Calvet}, {Gullbring}, \&
  {D'Alessio}}]{Hartmann98}
{Hartmann}, L., {Calvet}, N., {Gullbring}, E., \& {D'Alessio}, P. 1998, \apj,
  495, 385

\bibitem[{{Hogerheijde} \& {van der Tak}(2000)}]{Hogerheijde00}
{Hogerheijde}, M.~R., \& {van der Tak}, F.~F.~S. 2000, \aap, 362, 697

\bibitem[{{Jang-Condell}(2009)}]{Jang-Condell09}
{Jang-Condell}, H. 2009, \apj, 700, 820

\bibitem[{{Knez} {et~al.}(2009){Knez}, {Lacy}, {Evans}, {van Dishoeck}, \&
  {Richter}}]{Knez09}
{Knez}, C., {Lacy}, J.~H., {Evans}, N.~J., {van Dishoeck}, E.~F., \& {Richter},
  M.~J. 2009, \apj, 696, 471

\bibitem[{{Kurucz}(1993)}]{Kurucz93}
{Kurucz}, R.~L. 1993, VizieR Online Data Catalog, 6039, 0

\bibitem[{{Lada} {et~al.}(2006){Lada}, {Muench}, {Luhman}, {Allen}, {Hartmann},
  {Megeath}, {Myers}, {Fazio}, {Wood}, {Muzerolle}, {Rieke}, {Siegler}, \&
  {Young}}]{Lada06}
{Lada}, C.~J. {et~al.} 2006, \aj, 131, 1574

\bibitem[{{Lahuis} {et~al.}(2007){Lahuis}, {van Dishoeck}, {Blake}, {Evans},
  {Kessler-Silacci}, \& {Pontoppidan}}]{Lahuis07}
{Lahuis}, F., {van Dishoeck}, E.~F., {Blake}, G.~A., {Evans}, II, N.~J.,
  {Kessler-Silacci}, J.~E., \& {Pontoppidan}, K.~M. 2007, \apj, 665, 492

\bibitem[{{Lahuis} {et~al.}(2006){Lahuis}, {van Dishoeck}, {Boogert},
  {Pontoppidan}, {Blake}, {Dullemond}, {Evans}, {Hogerheijde}, {J{\o}rgensen},
  {Kessler-Silacci}, \& {Knez}}]{Lahuis06}
{Lahuis}, F. {et~al.} 2006, \apjl, 636, L145

\bibitem[{{Levison} \& {Morbidelli}(2003)}]{Levison03}
{Levison}, H.~F., \& {Morbidelli}, A. 2003, \nat, 426, 419

\bibitem[{{Lopez} {et~al.}(2008){Lopez}, {Antonelli}, {Wolf}, {Lagarde},
  {Jaffe}, {Navarro}, {Graser}, {Petrov}, {Weigelt}, {Bresson}, {Hofmann},
  {Beckman}, {Henning}, {Laun}, {Leinert}, {Kraus}, {Robbe-Dubois}, {Vakili},
  {Richichi}, {Abraham}, {Augereau}, {Behrend}, {Berio}, {Berruyer},
  {Chesneau}, {Clausse}, {Connot}, {Demyk}, {Danchi}, {Dugu{\'e}}, {Finger},
  {Flament}, {Glazenborg}, {Hannenburg}, {Heininger}, {Hugues}, {Hron},
  {Jankov}, {Kerschbaum}, {Kroes}, {Linz}, {Lizon}, {Mathias}, {Mathar},
  {Matter}, {Menut}, {Meisenheimer}, {Millour}, {Nardetto}, {Neumann},
  {Nussbaum}, {Niedzielski}, {Mosoni}, {Olofsson}, {Rabbia}, {Ratzka}, {Rigal},
  {Roussel}, {Schertl}, {Schmider}, {Stecklum}, {Thiebaut}, {Vannier}, {Valat},
  {Wagner}, \& {Waters}}]{Lopez08}
{Lopez}, B. {et~al.} 2008, in Presented at the Society of Photo-Optical
  Instrumentation Engineers (SPIE) Conference, Vol. 7013, Society of
  Photo-Optical Instrumentation Engineers (SPIE) Conference Series

\bibitem[{{Lyons} \& {Young}(2005)}]{Lyons05}
{Lyons}, J.~R., \& {Young}, E.~D. 2005, \nat, 435, 317

\bibitem[{{Machida} {et~al.}(2008){Machida}, {Kokubo}, {Inutsuka}, \&
  {Matsumoto}}]{Machida08}
{Machida}, M.~N., {Kokubo}, E., {Inutsuka}, S.-i., \& {Matsumoto}, T. 2008,
  \apj, 685, 1220

\bibitem[{{Meijerink} {et~al.}(2008){Meijerink}, {Poelman}, {Spaans},
  {Tielens}, \& {Glassgold}}]{Meijerink08}
{Meijerink}, R., {Poelman}, D.~R., {Spaans}, M., {Tielens}, A.~G.~G.~M., \&
  {Glassgold}, A.~E. 2008, \apjl, 689, L57

\bibitem[{Meijerink {et~al.}(2009)Meijerink, Pontoppidan, Blake, Poelman, \&
  Dullemond}]{Meijerink09}
Meijerink, R., Pontoppidan, K.~M., Blake, G.~A., Poelman, D., \& Dullemond,
  C.~P. 2009, \apj, submitted

\bibitem[{{Morbidelli} {et~al.}(2002){Morbidelli}, {Bottke}, {Froeschl{\'e}},
  \& {Michel}}]{Morbidelli02}
{Morbidelli}, A., {Bottke}, Jr., W.~F., {Froeschl{\'e}}, C., \& {Michel}, P.
  2002, Asteroids III, 409

\bibitem[{{Najita} {et~al.}(2003){Najita}, {Carr}, \& {Mathieu}}]{Najita03}
{Najita}, J., {Carr}, J.~S., \& {Mathieu}, R.~D. 2003, \apj, 589, 931

\bibitem[{{Najita} {et~al.}(2009){Najita}, {Doppmann}, {Bitner}, {Richter},
  {Lacy}, {Jaffe}, {Carr}, {Meijerink}, {Blake}, {Herczeg}, \&
  {Glassgold}}]{Najita09}
{Najita}, J.~R. {et~al.} 2009, \apj, 697, 957

\bibitem[{{Pascucci} {et~al.}(2007){Pascucci}, {Hollenbach}, {Najita},
  {Muzerolle}, {Gorti}, {Herczeg}, {Hillenbrand}, {Kim}, {Carpenter}, {Meyer},
  {Mamajek}, \& {Bouwman}}]{Pascucci07}
{Pascucci}, I. {et~al.} 2007, \apj, 663, 383

\bibitem[{{Pavlyuchenkov} {et~al.}(2007){Pavlyuchenkov}, {Semenov}, {Henning},
  {Guilloteau}, {Pi{\'e}tu}, {Launhardt}, \& {Dutrey}}]{Pavlyuchenkov07}
{Pavlyuchenkov}, Y., {Semenov}, D., {Henning}, T., {Guilloteau}, S.,
  {Pi{\'e}tu}, V., {Launhardt}, R., \& {Dutrey}, A. 2007, \apj, 669, 1262

\bibitem[{{Poelman} \& {Spaans}(2005)}]{Poelman05}
{Poelman}, D.~R., \& {Spaans}, M. 2005, \aap, 440, 559

\bibitem[{{Pontoppidan} {et~al.}(2008){Pontoppidan}, {Blake}, {van Dishoeck},
  {Smette}, {Ireland}, \& {Brown}}]{Pontoppidan08}
{Pontoppidan}, K.~M., {Blake}, G.~A., {van Dishoeck}, E.~F., {Smette}, A.,
  {Ireland}, M.~J., \& {Brown}, J. 2008, \apj, 684, 1323

\bibitem[{{Pontoppidan} {et~al.}(2005){Pontoppidan}, {Dullemond}, {van
  Dishoeck}, {Blake}, {Boogert}, {Evans}, {Kessler-Silacci}, \&
  {Lahuis}}]{Pontoppidan05}
{Pontoppidan}, K.~M., {Dullemond}, C.~P., {van Dishoeck}, E.~F., {Blake},
  G.~A., {Boogert}, A.~C.~A., {Evans}, II, N.~J., {Kessler-Silacci}, J.~E., \&
  {Lahuis}, F. 2005, \apj, 622, 463

\bibitem[{{Qi} {et~al.}(2008){Qi}, {Wilner}, {Aikawa}, {Blake}, \&
  {Hogerheijde}}]{Qi08}
{Qi}, C., {Wilner}, D.~J., {Aikawa}, Y., {Blake}, G.~A., \& {Hogerheijde},
  M.~R. 2008, \apj, 681, 1396

\bibitem[{{Rothman} {et~al.}(2005){Rothman}, {Jacquemart}, {Barbe}, {Chris
  Benner}, {Birk}, {Brown}, {Carleer}, {Chackerian}, {Chance}, {Coudert},
  {Dana}, {Devi}, {Flaud}, {Gamache}, {Goldman}, {Hartmann}, {Jucks}, {Maki},
  {Mandin}, {Massie}, {Orphal}, {Perrin}, {Rinsland}, {Smith}, {Tennyson},
  {Tolchenov}, {Toth}, {Vander Auwera}, {Varanasi}, \& {Wagner}}]{Rothman05}
{Rothman}, L.~S. {et~al.} 2005, Journal of Quantitative Spectroscopy and
  Radiative Transfer, 96, 139

\bibitem[{{Sabatke} {et~al.}(2005){Sabatke}, {Burge}, \& {Sabatke}}]{Sabatke05}
{Sabatke}, E., {Burge}, J., \& {Sabatke}, D. 2005, \ao, 44, 1360

\bibitem[{{Salyk} {et~al.}(2007){Salyk}, {Blake}, {Boogert}, \&
  {Brown}}]{Salyk07}
{Salyk}, C., {Blake}, G.~A., {Boogert}, A.~C.~A., \& {Brown}, J.~M. 2007,
  \apjl, 655, L105

\bibitem[{{Salyk} {et~al.}(2008){Salyk}, {Pontoppidan}, {Blake}, {Lahuis}, {van
  Dishoeck}, \& {Evans}}]{Salyk08}
{Salyk}, C., {Pontoppidan}, K.~M., {Blake}, G.~A., {Lahuis}, F., {van
  Dishoeck}, E.~F., \& {Evans}, II, N.~J. 2008, \apjl, 676, L49

\bibitem[{{Sch{\"o}ier} {et~al.}(2005){Sch{\"o}ier}, {van der Tak}, {van
  Dishoeck}, \& {Black}}]{Schoier05}
{Sch{\"o}ier}, F.~L., {van der Tak}, F.~F.~S., {van Dishoeck}, E.~F., \&
  {Black}, J.~H. 2005, \aap, 432, 369

\bibitem[{{Semenov} {et~al.}(2008){Semenov}, {Pavlyuchenkov}, {Henning},
  {Wolf}, \& {Launhardt}}]{Semenov08}
{Semenov}, D., {Pavlyuchenkov}, Y., {Henning}, T., {Wolf}, S., \& {Launhardt},
  R. 2008, \apjl, 673, L195

\bibitem[{{Siess} {et~al.}(2000){Siess}, {Dufour}, \& {Forestini}}]{Siess00}
{Siess}, L., {Dufour}, E., \& {Forestini}, M. 2000, \aap, 358, 593

\bibitem[{Smith {et~al.}(2009)Smith, Pontoppidan, Young, Morris, \& van
  Dishoeck}]{Smith09}
Smith, R., Pontoppidan, K.~M., Young, E.~D., Morris, M.~R., \& van Dishoeck,
  E.~F. 2009, \apj, in press

\bibitem[{{Strom} {et~al.}(1989){Strom}, {Strom}, {Edwards}, {Cabrit}, \&
  {Skrutskie}}]{Strom89}
{Strom}, K.~M., {Strom}, S.~E., {Edwards}, S., {Cabrit}, S., \& {Skrutskie},
  M.~F. 1989, \aj, 97, 1451

\bibitem[{{Thi} \& {Bik}(2005)}]{Thi05}
{Thi}, W.-F., \& {Bik}, A. 2005, \aap, 438, 557

\bibitem[{{Tsiganis} {et~al.}(2005){Tsiganis}, {Gomes}, {Morbidelli}, \&
  {Levison}}]{Tsiganis05}
{Tsiganis}, K., {Gomes}, R., {Morbidelli}, A., \& {Levison}, H.~F. 2005, \nat,
  435, 459

\bibitem[{{Udry} \& {Santos}(2007)}]{Udry07}
{Udry}, S., \& {Santos}, N.~C. 2007, \araa, 45, 397

\bibitem[{{van Boekel} {et~al.}(2009){van Boekel}, {G{\"u}del}, {Henning},
  {Lahuis}, \& {Pantin}}]{vanBoekel09}
{van Boekel}, R., {G{\"u}del}, M., {Henning}, T., {Lahuis}, F., \& {Pantin}, E.
  2009, \aap, 497, 137

\bibitem[{{van der Plas} {et~al.}(2009){van der Plas}, {van den Ancker},
  {Acke}, {Carmona}, {Dominik}, {Fedele}, \& {Waters}}]{vanderplas09}
{van der Plas}, G., {van den Ancker}, M.~E., {Acke}, B., {Carmona}, A.,
  {Dominik}, C., {Fedele}, D., \& {Waters}, L.~B.~F.~M. 2009, \aap, 500, 1137

\bibitem[{{van der Plas} {et~al.}(2008){van der Plas}, {van den Ancker},
  {Fedele}, {Acke}, {Dominik}, {Waters}, \& {Bouwman}}]{vanderplas08}
{van der Plas}, G., {van den Ancker}, M.~E., {Fedele}, D., {Acke}, B.,
  {Dominik}, C., {Waters}, L.~B.~F.~M., \& {Bouwman}, J. 2008, \aap, 485, 487

\bibitem[{{van Zadelhoff} {et~al.}(2001){van Zadelhoff}, {van Dishoeck}, {Thi},
  \& {Blake}}]{vanZadelhoff01}
{van Zadelhoff}, G.-J., {van Dishoeck}, E.~F., {Thi}, W.-F., \& {Blake}, G.~A.
  2001, \aap, 377, 566

\bibitem[{{Willacy}(2007)}]{Willacy07}
{Willacy}, K. 2007, \apj, 660, 441

\bibitem[{{Woitke} {et~al.}(2009){Woitke}, {Kamp}, \& {Thi}}]{Woitke09}
{Woitke}, P., {Kamp}, I., \& {Thi}, W.-F. 2009, ArXiv e-prints

\bibitem[{{Wolf} \& {D'Angelo}(2005)}]{Wolf05}
{Wolf}, S., \& {D'Angelo}, G. 2005, \apj, 619, 1114

\bibitem[{{Wooden} {et~al.}(2007){Wooden}, {Desch}, {Harker}, {Gail}, \&
  {Keller}}]{WoodenPPV}
{Wooden}, D., {Desch}, S., {Harker}, D., {Gail}, H.-P., \& {Keller}, L. 2007,
  in Protostars and Planets V, ed. B.~{Reipurth}, D.~{Jewitt}, \& K.~{Keil},
  815--833

\bibitem[{{Woods} \& {Willacy}(2009)}]{Woods09}
{Woods}, P.~M., \& {Willacy}, K. 2009, \apj, 693, 1360

\end{thebibliography}

\end{document}